%% file: paper.tex
\makeatletter
\newcommand\footnoteref[1]{\protected@xdef\@thefnmark{\ref{#1}}\@footnotemark}
\makeatother

\documentclass[table,xcdraw,sigconf]{acmart-modified}

\setcopyright{iw3c2w3}

\usepackage[normalem]{ulem}
\useunder{\uline}{\ul}{}

\definecolor{my-yellow}{HTML}{ffff00}
\definecolor{my-green}{HTML}{00ff00}
\definecolor{my-red}{HTML}{ff0000}

\usepackage{booktabs} 
\usepackage{hyperref}
\usepackage{algorithm}
\usepackage{algpseudocode}
\usepackage{pdfpages}
\usepackage{changepage}
\usepackage{caption}
\usepackage{microtype} 

\usepackage[skip=7pt]{caption}
\usepackage{multirow} 
\usepackage[most]{tcolorbox} 

\usepackage[bottom]{footmisc}

\usepackage{amsmath}
\usepackage{graphicx}
\usepackage{xcolor}
\usepackage{multicol}
\usepackage{tikz}

\definecolor{light-yellow}{RGB}{255, 255, 204}
\newtcolorbox[]{summary}[1][]{colframe=black!45, boxrule=.3mm, 
    colback=light-yellow,
    boxsep=.25mm,
    left=1.5mm,
    right=1.5mm,
  ,#1}

\usepackage{siunitx} 
\usepackage{listings} 
\newcommand{\new}[1]{\textcolor{black}{#1}}

\newcommand{\reviewerfix}[1]{\textcolor{black}{#1}}

\algnewcommand{\LineComment}[1]{\State \(\triangleright\) #1}

\makeatother

\begin{document}

    \makeatletter
    \@namedef{r@tocindent4}{30pt}
    \@namedef{r@tocindent5}{30pt}
    \makeatother
    \copyrightyear{2021} 
    \acmYear{2021} 
    \settopmatter{printacmref=false}
      \title[Spatial Indexing for Accelerated Feature Retrieval]{\huge{Exploring Spatial Indexing for Accelerated Feature Retrieval in HPC}}

    \author{Margaret Lawson}
    \affiliation{
      \institution{The University of Illinois at Urbana-Champaign}
    \country{}
    }
    \additionalaffiliation{
      \institution{Sandia National Laboratories}
      \city{Albuquerque}
      \state{New Mexico}
        \country{}
    }
    \email{mlawson4@illinois.edu}
    
    \author{William Gropp}
    \affiliation{
      \institution{The University of Illinois at Urbana-Champaign}
        \country{}
    }
    \email{wgropp@illinois.edu}
    
    \author{Jay Lofstead}
    \affiliation{
      \institution{Sandia National Laboratories}
      \city{Albuquerque}
      \state{New Mexico}
        \country{}
    }
    \email{gflofst@sandia.gov}

    \input{abstract}

    \keywords{geometric range searching, spatial indexing, k-d tree, R-tree, octree}

\maketitle

\input{intro}

\input{related-work}
\input{libraries-overview}

\input{eval}
\input{discussion-future-work}
\input{acknowledgements}

  \bibliographystyle{ACM-Reference-Format}
  \bibliography{bibliography}

\end{document}

%% file: abstract.tex
\begin{abstract}
    Despite the critical role that range queries play in analysis and visualization for HPC applications, there has been no comprehensive analysis of indices that are designed to accelerate range queries and the extent to which they are viable in an HPC setting. In this state of the practice paper we present the first such evaluation, examining 20 open-source C and C++ libraries that support range queries. Contributions of this paper include answering the following questions: which of the implementations are viable in an HPC setting, how do these libraries compare in terms of build time, query time, memory usage, and scalability, what are other trade-offs between these implementations, is there a single overall best solution, and when does a brute force solution offer the best performance? We also share key insights learned during this process that can assist both HPC application scientists and spatial index developers.
\end{abstract}

%% file: intro.tex
\section{Introduction}
\label{sec:intro}
During analysis and visualization, HPC application scientists often need to extract particular spatial subsets of their data. Scientists may need to extract these subsets for sub-sampling~\cite{Lofstead:2011:six-degrees}, to perform areal interpolation~\cite{papadomanolakis:2006:query-processing-unstructured-tet-meshes, hinterberger:1994:region-data-interpolation}, or to perform spatial correlations~\cite{lu:2017:region-based-data-correlation}. Scientists may also extract spatial subsets that correspond to regions or features of interest. This can allow scientists to concentrate their analysis and visualization on more interesting data or to use different analysis routines depending on the type of region. Examples of regions that may be extracted include application boundaries and surfaces~\cite{papadomanolakis:2006:querying-unstructured-tetrahedral-meshes}, 
regions of the brain in neuroscience imagery~\cite{beyer:2013:viz-region-of-interest-neuroscience, tauheed:2012:range-queries-brain-simulations}, counties or other regional boundaries in satellite imagery~\cite{beynon:2000:analysis-viz-region-of-interest}, regions with different flow types in a fluid dynamics simulation~\cite{barone:2021:fluid-flow-classification}, and regions with different chemical properties in a combustion simulation~\cite{gosink:2007:variable-interactions-query-driven-viz}. 
Examples of spatial features of interest that may be extracted include supernovae in astronomy~\cite{betoule:2013:sdss-supernovae}, blood vessels, lesions, and areas of inflammation in pathology imagery~\cite{aji:2012:spatial-query-medical-segmentation-region-of-interest, liang:2016:spatial-queries-pathology-mapreduce, liang:2017:ispeed-spatial-querying}, tissue regions with particular cell types in microscopy datasets~\cite{beynon:2000:analysis-viz-region-of-interest}, shocks in a fluid simulation~\cite{beynon:2000:analysis-viz-region-of-interest}, flame fronts, extinction events and vortices in a combustion simulation~\cite{gosink:2007:variable-interactions-query-driven-viz, gosink:2010:viz-region-of-interest}, and tropical storms in a climate simulation~\cite{chiu:2015:analysis-region-of-interest}. 
    These regions and features may be identified using techniques such as image segmentation~\cite{pecha:2016:viz-segmentation-region-of-interest, pradipkumar:2016:viz-segmentation-medical}, machine-learning based approaches~\cite{graff:2014:neural-nets-astronomy, liu:2016:neural-net-climate, tzeng:2005:intelligent-feature-extraction, cubuk:2015:machine-learning-physics}, or other analysis techniques~\cite{bremer:2010:interactive-exploration-topology-segmentation, stockinger:2005:query-driven-viz, post:2003:flow-visualisation, wang:2008:importance-driven-viz, livnat:1996:isosurface-extraction, yu:2010:in-situ-viz-combustion}. Extracting spatial subsets is thus a critical component of many different types of analysis and visualization routines for a wide range of HPC applications.

Extracting a particular spatial subset is trivial for regular meshes (meshes in which the cells are congruent parallelotopes) where a single, simple function can be used to map from a spatial coordinate to a data array offset. In the simplest case (a Cartesian mesh), the mesh coordinates will be the same as the array indices. In contrast, for unstructured meshes (meshes without implied neighborhood connectivity), this operation is very computationally intensive. Without any form of index, scientists will have to perform a linear (brute force) search over all mesh points or elements for each region they want to extract. With petascale simulations already using meshes that can range from tens of billions~\cite{rasquin:2014:mesh-tens-billion-elements, wang:2017:mesh-tens-billions-elements, houzeaux:2011:mesh-tens-billions-elements, zhao:2020:mesh-tens-billions-elements} to trillions~\cite{godenschwager:2013:trillions-elements} of elements and with the impending arrival of exascale machines~\cite{doe:2019:first-exascale-aurora}, this kind of linear search, which may need to be performed thousands or millions of times during analysis and visualization, represents a significant inefficiency that can delay the path to scientific discovery. A lot of research has been done to develop spatial indices that can accelerate this type of search (known as a range query) to help scientists quickly identify what mesh points or elements fall within a given region. However, there has been no comprehensive analysis of spatial index implementations that are designed to accelerate range queries and the extent to which they are viable in an HPC setting. 
    \new{To be viable in an HPC setting, a spatial indexing library must have good performance for building the index and performing queries, good scalability, and moderate memory overheads. It is important for the index to have moderate memory overheads since the index will provide the best performance when stored in memory, many applications require significant amounts of memory (for MPI, for the mesh, for variable data, etc.) and nodes have limited amounts of memory.} 

In this \reviewerfix{state of the practice} paper, we present the first comprehensive review of spatial index implementations that support range queries. We also provide scientists with the information they need to decide when to use a spatial index for range querying and which spatial index implementation will be best suited to their use case. This information can help scientists accelerate range queries by several orders of magnitude and thereby greatly accelerate their analysis routines and the discovery process. Contributions of this paper include presenting a thorough analysis of 20 free, open-source C and C++ libraries that support range queries with expected sub-linear query times, \reviewerfix{sharing key insights learned during this process,} and answering the following questions:
    \begin{enumerate}
        \item Which of the implementations are viable in an HPC setting?
        \item How do these libraries compare in terms of build time, query time, and memory usage at different scales? 
        \item What are other factors in deciding which library to use?
        \item Is there a single overall best solution?
        \item When does a brute force solution offer the best performance?
    \end{enumerate}

The rest of this paper is organized as follows. Section~\ref{sec:related-work} discusses related work. 
    Section~\ref{sec:libraries-overview} provides an overview of the evaluated libraries. Section~\ref{sec:eval} presents the evaluation and results. 
Finally, Section~\ref{sec:future-work} provides discussion, \reviewerfix{offers additional insights, and} presents areas for future work.

%% file: related-work.tex
\section{Related Work}
\label{sec:related-work}
    Range queries have been extensively studied in many different branches of computer science including privacy and security~\cite{zhu:2015:range-queries-privacy, xu:2018:range-queries-encryption}, mobile computing~\cite{cong:2016:mobile-computing-range-queries}, networks and other graph-based areas~\cite{demirbas:2003:range-queries-sensor-networks, soheili:2005:range-quereies-sensor-networks, wan:2019:range-queries-voronoi-diagram-IOT}, cloud computing~\cite{yu:2015:geospark, tang:2016:locationspark, eldawy:2015:spatialhadoop, zhong:2012:range-queries-mapreduce}, and databases~\cite{papadias:2003:spatial-network-databases-range-queries, shekhar:1999:spatial-databases, egenhofer:1994:spatial-SQL-databases, orenstein:1986:spatial-databases-range-queries, guting:1994:spatial-databases}. Below we provide an overview of two categories of related work that we find to be most related to the work presented here: theoretical research done on geometric range searching and search structures and research that either directly implements structures that support range queries or evaluates these structures.

\subsection{Theory: Geometric Range Searching and Geometric Search Structures}
Efficient geometric range searching has been the focus of a significant amount of research in the theoretical branch of computer science
    ~\cite{matouvsek:1994:geometric-range-searching, agarwal:1996:range-searching, willard:1985:data-structures-orthogonal-range-queries,  chazelle:1990:lower-bounds-orthogonal-range-searching-part1, hadjieleftheriou:2005:theory-range-queries, jagadish:1990:theory-range-queries-polyhedra, pagel:1993:theory-range-queries-structures-analysis,  bentley:1980:theory-performance-bounds-range-query-structures,  overmars:1988:theory-range-queries-grid-structures-analysis}. 
   This work has invented a number of geometric search structures, and here we present an overview of the three that are most commonly supported by the libraries we evaluate in this paper: octrees, k-d trees, and R-trees.

\subsubsection{Octrees}
    An octree~\cite{jackins:1980:octree-first} is a tree in which each internal node has exactly eight children and which is not necessarily balanced. An octree can either be used to recursively partition a 3D space into eight octants (region-based octrees) or to partition a set of 3D points based on their coordinates (point-based octrees). 
    Internally, octrees typically store point data only in the leaf nodes for region-based octrees and typically store point data both at leaf nodes and internal nodes for point-based octrees~\cite{formaggia:1999:data-structures-geometric-search}.

\sloppy
\subsubsection{K-d trees}
\sloppy
\sloppy
        K-d trees~\cite{bentley:1975:kdtree-first} are binary trees that store k-\linebreak dimensional data and represent a recursive subdivision of this data using $(k-1)$-dimensional hyperplanes. Each (non-leaf) level in the tree corresponds to one of the k-dimensions, and each internal node at that level represents a splitting hyperplane for that dimension. K-trees are sensitive to the order in which points are inserted and are not necessarily balanced. Typically the data will be split before the median point along the longest side of the node (the sliding midpoint rule), but other splitting strategies are possible.

\fussy

\subsubsection{R-trees}
    R-trees~\cite{guttman:1984:rtree-first} are balanced trees that contain a hierarchy of d-dimensional boxes, where each child node is contained in the box represented by its parent. Depending on the implementation, these partitions may overlap. The leaves either store a d-dimensional point or the d-dimensional minimum bounding box of the objects stored in the leaf. These objects may be a set of points or other shapes. R-trees have a minimum and maximum branching factor for all internal nodes (apart from the root), and there are algorithms designed to build optimal R-trees using bulk (static) construction~\cite{leutenegger:1997:rtree-packing-str, garcia:1998:greedy-alg-bulk-loading-rtree}. R*-trees are an R-tree variant that is designed to minimize coverage and spatial overlap between internal nodes.

\subsection{Implementations and Evaluations of Structures Supporting Range Queries}
    Several recent research efforts have implemented data structures that support range queries. However, to our knowledge, none of these implementations are open-source. This includes the development of a query processing technique for unstructured tetrahedral meshes in PostgreSQL~\cite{papadomanolakis:2006:query-processing-unstructured-tet-meshes}, intersection queries in PostGIS~\cite{de-olivera:2020:query-3d-spatial-database}, point, range and polygon queries in Microsoft's SQL Server~\cite{perlman:2010:organization-data-non-convex-spatial-domains}, and parallel range queries using GPUs~\cite{nguyen:2016:range-queries-unstructured-meshes, nouri:2018:range-queries-gpu, kim:2013:range-queries-gpu-rtree, prasad:2015:rtree-gpu, luo:2012:rtree-gpu, you:2013:rtree-gpu-bulk-loading}. 
    One project performed a comparison of nearest neighbor search implementations and included some evaluation of range queries~\cite{elseberg:2012:libnabo-comparison-knn-algs-shape-registration}. However, this work only considers six of the libraries we evaluate here, collects range query results only for artificial datasets containing 60,000 points, and was performed almost a decade ago. 

\subsection{Summary}
\new{Our work differs from related work in that it focuses on spatial index implementations that are open-source and which support range queries. Most related work is focused on structures that are theoretical rather than implemented, structures that do not offer range queries, or structures that offer range queries but are not open-source. This work also differs from related work in that it evaluates structures for use in HPC and therefore uses significantly larger datasets than those used in related work and offers an analysis of memory requirements, and strong and weak scaling.}

%% file: libraries-overview.tex
\section{Spatial Indexing Libraries Overview}
\label{sec:libraries-overview}

        We evaluate 20 libraries that support range queries. To be included in this study, a library has to: provide bindings for C or C++, offer a free, open-source version, and offer spatial indexing that supports range querying (with expected sub-linear query times). We limit our search to libraries with C or C++ bindings since these are the languages most commonly used in HPC \reviewerfix{for analysis and visualization tasks~\cite{amaral:2020:hpc-programming-languages, laguna:2019:mpi-usage-hpc}}. Only one library, LEDA~\cite{mehlhorn:1999:leda}, is excluded for failing to meet the free, open-source requirement. LEDA offers a free version, but range queries are only offered in its paid version, which costs 3000 euros for pure research efforts. 
        \reviewerfix{We leave for future work evaluating libraries that offer other language bindings.}

    Table~\ref{tab:basic-info-libraries} contains basic information about the libraries. 
   In the evaluation, we perform box queries since this is the type of query used most commonly in analysis and visualization. For libraries that do not support box queries, we use a sphere (or radius) that is large enough to encompass the box, and then perform a filtering step to ensure exact matches are returned. 
   This will result in a slight performance penalty for these libraries.

\input{tables/libs-summary-single-column}

Note that some of the libraries share the same name or similar names (e.g., KDTREE and KDTREE2, libkdtree++ and libkdtree). We have added a number (shown in parentheses) after some of these library names to help disambiguate them. For the rest of the paper, we will refer to these libraries using the name appended with the number (without the parentheses). 
    \begin{summary}
        Insight: few libraries can store mesh elements (shapes/boxes). Since many variables are calculated per mesh element, this is a significant limitation.
    \end{summary}

    For more information about these libraries, readers can look at 
    the Git repository for this project~\cite{lawson:2021:range_tree_benchmarking_suite_git_repo}.
    \vspace{-.02in}

%% file: tables/libs-summary-single-column.tex
\begin{table}[!htbp]
\vspace{-.1in}
\begin{adjustwidth}{-.5cm}{}
\small
\setlength{\tabcolsep}{2.5pt}
\centering
\caption{Basic info. about the libraries evaluated in this paper}
\label{tab:basic-info-libraries}
\begin{tabular}{|l|c|c|c|c|c|} 
\hline
\multicolumn{1}{|c|}{\textbf{Library}} & \multicolumn{1}{c|}{\textbf{Ver.}} & \multicolumn{1}{c|}{\begin{tabular}[c]{@{}c@{}}\textbf{Tree}\\\textbf{type}\end{tabular}} & \multicolumn{1}{c|}{\begin{tabular}[c]{@{}c@{}}\textbf{Search}\\\textbf{type}\end{tabular}} & \multicolumn{1}{c|}{\begin{tabular}[c]{@{}c@{}}\textbf{Data}\\\textbf{types}\end{tabular}} & \multicolumn{1}{c|}{\begin{tabular}[c]{@{}c@{}}\textbf{\small Can set}\\\textbf{\small leaf size}\end{tabular}}  \\ 
\hline
\textbf{3DTK~\cite{nuchter:2011:3dtk}}                          & \small -                                     & \textbf{k-d tree}                                                                                & \textbf{Box}                                                                              & \textbf{Points}                                                                                     & \textbf{Yes}                                                                                          \\ 
\hline
\textbf{ALGLIB~\cite{bochkanov:2011:alglib}}                       & \textbf{\small 3.17.0}                       & \textbf{k-d tree}                                                                                & \textbf{Box}                                                                              & \textbf{Points}                                                                                     & \textbf{No}                                                                                           \\ 
\hline
\textbf{ANN~\cite{arya:1998:ann}}                           & \textbf{\small 1.1.2}                        & \textbf{k-d tree}                                                                                & \begin{tabular}[c]{@{}c@{}}\textbf{KNN w.}\\\textbf{radius}\end{tabular}                                                                      & \textbf{Points}                                                                                     & \textbf{Yes}                                                                                          \\ 

\hline
\textbf{ANN~\cite{arya:1998:ann}}                           & \textbf{\small 1.1.2}                        & \textbf{BD-tree}                                                                                 & \begin{tabular}[c]{@{}c@{}}\textbf{KNN w.}\\\textbf{radius}\end{tabular}                                                                      & \textbf{Points}                                                                                     & \textbf{Yes}                                                                                          \\ 
\hline
\textbf{Boost~\cite{boost-geometry-website}}                & \textbf{\small 1.74.0}                       & \textbf{R-tree}                                                                                  & \textbf{Box}                                                                              & \textbf{Points, boxes}                                                                               & \textbf{Yes}                                                                                          \\ 
\hline
\textbf{CGAL~\cite{fabri:2009:cgal}}                          & \textbf{\small 5.2.0}                        & \textbf{k-d tree}                                                                                & \textbf{Box}                                                                              & \textbf{Points}                                                                                     & \textbf{Yes}                                                                                          \\ 
\hline
\textbf{CGAL~\cite{fabri:2009:cgal}}                          & \textbf{\small 5.2.0}                        & \textbf{Box}                                                                              & \textbf{Box}                                                                              & \textbf{Points}                                                                                     & \textbf{No}                                                                                           \\ 
\hline
\textbf{CGAL~\cite{fabri:2009:cgal}}                          & \textbf{\small 5.2.0}                        & \textbf{Segment}                                                                            & \textbf{Box}                                                                              & \textbf{Boxes}                                                                                      & \textbf{No}                                                                                           \\ 
\hline
\textbf{CGAL~\cite{fabri:2009:cgal}}                          & \textbf{\small 5.2.0}                        & \textbf{R-tree}                                                                                 & \textbf{Box}                                                                              & \textbf{Points, shapes}                                                                            & \textbf{No}                                                                                           \\ 
\hline
\textbf{FLANN~\cite{muja:2009:flann}}                         & \textbf{\small 1.9.1}                        & \textbf{k-d tree}                                                                                & \textbf{Sphere}                                                                             & \textbf{Points}                                                                                     & \textbf{Yes}                                                                                          \\ 
\hline
\textbf{KDTREE(1)~\cite{kdtree-git}}                     & \small -                                     & \textbf{k-d tree}                                                                                & \textbf{Sphere}                                                                             & \textbf{Points}                                                                                     & \textbf{No}                                                                                           \\ 
\hline
\textbf{KDTREE2~\cite{kennel:2004:kdtree2}}                       & \small -                                     & \textbf{k-d tree}                                                                                & \textbf{Sphere}                                                                             & \textbf{Points}                                                                                     & \textbf{Yes}                                                                                          \\ 
\hline
\textbf{KDTREE(3)~\cite{kdtree3-git}}                     & \small -                                     & \textbf{k-d tree}                                                                                & \textbf{Sphere}                                                                             & \textbf{Points}                                                                                     & \textbf{Yes}                                                                                          \\ 
\hline
\textbf{KDTREE(4)~\cite{kdtree4-git}}                     & \textbf{\small 0.5.7}                        & \textbf{k-d tree}                                                                                & \textbf{Sphere}                                                                             & \textbf{Points}                                                                                     & \textbf{No}                                                                                           \\ 
\hline
\textbf{libkdtree++~\cite{libkdtree-git}}                   & \small -                                     & \textbf{k-d tree}                                                                                & \textbf{Sphere}                                                                             & \textbf{Points}                                                                                     & \textbf{No}                                                                                           \\ 
\hline
\textbf{libkdtree(2)~\cite{libkdtree2-git}}                  & \small -                                     & \textbf{k-d tree}                                                                                & \textbf{Box}                                                                              & \textbf{Points}                                                                                     & \textbf{No}                                                                                           \\ 
\hline
\textbf{libnabo~\cite{elseberg:2012:libnabo-comparison-knn-algs-shape-registration}}                       & \textbf{\small 1.0.7}                        & \textbf{k-d tree}                                                                                & \begin{tabular}[c]{@{}c@{}}\textbf{KNN w.}\\\textbf{radius}\end{tabular}                                                                      & \textbf{Points}                                                                                     & \textbf{Yes}                                                                                          \\ 
\hline
\textbf{libspatialindex~\cite{hadjieleftheriou:2015:libspatialindex}}                & \textbf{\small 1.9.3}                        & \textbf{R*-tree}                                                                                 & \textbf{Box}                                                                              & \textbf{Boxes}                                                                                      & \textbf{Yes}                                                                                          \\ 
\hline
\textbf{nanoflann~\cite{blanco:2014:nanoflann}}                     & \textbf{1.3.2}                        & \textbf{k-d tree}                                                                                & \textbf{Sphere}                                                                             & \textbf{Points}                                                                                     & \textbf{Yes}                                                                                          \\ 
\hline
\textbf{Octree~\cite{behley:2015:octree-library}}                        & \small -                                     & \textbf{octree}                                                                                  & \textbf{Sphere}                                                                             & \textbf{Points}                                                                                     & \textbf{Yes}                                                                                          \\ 
\hline
\textbf{PCL~\cite{rusu:2011:pcl}}                           & \textbf{\small 1.11.1}                       & \textbf{k-d tree}                                                                               & \textbf{Sphere}                                                                             & \textbf{Points}                                                                                     & \textbf{No}                                                                                           \\ 
\hline
\textbf{PCL~\cite{rusu:2011:pcl}}                           & \textbf{\small 1.11.1}                       & \textbf{octree}                                                                                  & \textbf{Box}                                                                              & \textbf{Points}                                                                                     & \textbf{Yes}                                                                                          \\ 
\hline
\textbf{PicoTree~\cite{pico-tree-git}}                    & \textbf{\small 0.5.2}                        & \textbf{k-d tree}                                                                                & \textbf{Box}                                                                              & \textbf{Points}                                                                                     & \textbf{Yes}                                                                                          \\ 
\hline
\textbf{R-tree~\cite{rtree-template-website}}               & \small -                                     & \textbf{R-tree}                                                                                  & \textbf{Box}                                                                              & \textbf{Boxes}                                                                                      & \textbf{Yes}                                                                                          \\ 
\hline
\textbf{Spatial~\cite{spatial-website}}                       & \textbf{\small 2.1.8}                        & \textbf{k-d tree}                                                                                & \textbf{Box}                                                                              & \textbf{Points, boxes}                                                                                & \textbf{No}                                                                                           \\
\hline
\end{tabular}
\end{adjustwidth}
\vspace{-.2in}
\end{table}

%% file: eval.tex
\section{Evaluation and Results}
\label{sec:eval}
    This section presents an overview of the evaluation setup including the hardware and software, datasets, and scales used in testing. We also present the evaluation results for storage and querying of mesh points and mesh elements at two different scales, and present an analysis of the scalability of the libraries. \reviewerfix{HPC application data is typically collected for each mesh node or mesh element. This is why we perform an evaluation for both mesh points and elements.}

\vspace{-.03in}

\subsection{Experimental Setup}
    All experiments are run on the Vortex machine at Sandia National Labs. Vortex uses RHEL7, and has the following per node: 318 GB high-bandwidth DRAM and dual socket IBM POWER9 CPUs with 22 cores/socket and 4 hardware threads per core (176 total). We use GCC 10.2.0 to compile all of the libraries apart from PCL, which we can only build with GCC 8.3.1 on Vortex. 
    We use four different evaluation setups, which are described in Table~\ref{tab:eval-configs}. All experiments use the same mesh, which was created for a typical~\cite{domino:2019:large-unstructured-meshes} turbulent low-Mach study that includes solution verification. The turbulent low-Mach study uses meshes that range from around 150 million to 2 billion nodes and elements. The mesh we use is unstructured and has 152.7 million nodes and 152.1 million hexahedral elements. The coordinates are 8 byte doubles. It should be noted that there is no standard in HPC for the number of mesh points or elements assigned per process during domain decomposition. The number will depend on a number of factors such as the computational intensity of the application, the memory per node, and the number of compute hours the scientist has access to. However, using between $10,000$ and $100,000$ nodes or elements per process during a simulation is common\reviewerfix{~\cite{gahvari:2011:10k-to-100k-mesh-pts-per-process, hsu:2011:10k-to-100k-mesh-pts-per-process, quintanas:2018:10k-to-100k-mesh-pts-per-process, gabriel:2010:10k-to-100k-mesh-pts-per-process, feki:2008:10k-to-100k-mesh-pts-per-process, zaspel:2011:10k-to-100k-mesh-pts-per-process, colli:2018:10k-to-100k-mesh-pts-per-process, garcia:2020:10k-to-100k-mesh-pts-per-process, landge:2014:in-situ-merge-trees}}. Then, during analysis and visualization, scientists will often use a much smaller number of processors (10\% or fewer)\reviewerfix{~\cite{marrinan:2019:procs-simulation-vs-analysis, pena:2021:procs-simulation-vs-analysis, cummings:2008:procs-simulation-vs-analysis}} thus resulting in upwards of $100,000$ to $1,000,000$ mesh nodes or elements per process. \reviewerfix{This is further evidenced by the fact that analysis and visualization clusters often have far fewer cores and nodes than compute clusters~\cite{simakov:2018:analysis-viz-clusters, salim:2019:analysis-viz-clusters}.} We therefore perform ``small'' scale evaluation with approximately $1,000,000$ mesh nodes or elements per process to reflect this common range, and ``large'' scale evaluation with approximately $10,000,000$ mesh nodes or elements per process to reflect more extreme use cases. We use OpenMPI 4.1.0 with one hardware thread per (MPI) process. Apart from the weak scaling results, all evaluations use a single node. But, as the weak scaling results demonstrate in Section~\ref{sec:weak-scaling}, these results should hold at any scale if the number of processes per node is held constant. \reviewerfix{This is because each process performs entirely independent work without any need for message passing or coordination, and the indices are also kept entirely in-memory (eliminating the possibility of file system contention).} 
    
\begin{table}[!htbp]
\small
\vspace{-.1in}
\caption{Experiment setups}
\label{tab:eval-configs}
\begin{tabular}{|l|l|l|}
\hline & \textbf{Mesh points} & \textbf{Mesh elements} \\ \hline
\textbf{ 160 procs. (small scale)} & {1,227,411 / proc.} & {1,087,807 / proc.}  \\ \hline
\textbf{ 16 procs. (large scale)} & {10,163,37 / proc.} & {10,878,078 / proc.} \\ \hline
\end{tabular}
\vspace{-.15in}
\end{table}

\subsection{Performance Evaluations}\label{sec:performance-evals}
Each performance evaluation consists of two basic parts. 
First, each process creates an in-memory instance of the data structure being evaluated and inserts the mesh coordinates or elements for its subdomain (assigned portion of the simulation space).  
Second, each process performs a set of queries at four different sizes, which we refer to as extra-small, small, medium and large. 
These queries return approximately 0.001\%, 0.1\% 1\%, and 10\% of the data. Each process performs 10,000 queries that are extra-small, 10,000 that are small, 10,000 that are medium, and 1,000 that are large. This allows us to evaluate how the spatial indices perform for retrieving features or regions of interest that range from very small to very large. 
\reviewerfix{This testing workload is designed to reflect the real-world scenario described and cited in the introduction: when a scientist performing analysis or visualization with a complex mesh type needs to extract data for particular regions of interest.}
In the large scale evaluation, we evaluate all libraries that perform within a factor of 10$\times$ of the best performing library for each of the query sizes. 
This helps us to identify which libraries perform ``best'' across different query sizes and at different scales. 
For all libraries that allow the user to set a maximum leaf size, we test leaf sizes of 20 and 200. 
\reviewerfix{For space reasons, we only present the best result for each library.} 
The full set of results can be found in the Git repository for this project~\cite{lawson:2021:range_tree_benchmarking_suite_git_repo}. Several libraries also offer different insertion algorithms or splitting strategies (for k-d or BD trees). However, we did not find any significant performance differences between these algorithms and have therefore omitted these results to save space. One exception is that we use Boost's STR packing algorithm for all of the Boost tests since it results in dramatically improved results for data that can be bulk loaded (like mesh data). For libraries that offer both static and dynamic trees, we use static trees since the mesh we use in evaluation is static. 
We perform three runs for each configuration and average the results. 
After completing the timing results, we also collect memory results using the Massif heap profiler~\cite{valgrind-massif-manual} from Valgrind 3.16.1~\cite{nethercote:2007:valgrind}. For these memory results, we present data only on the construction of the tree since memory increases during querying are transient. 
We exclude memory allocated as a result of reading in the mesh nodes or elements from the mesh file. 
We run these tests on a single process that has the amount of data closest to the average (within 3\% of the global average). 

In addition to the libraries, we evaluate a brute force solution. In the build phase, the solution copies the 2D vector of mesh points or elements using the assignment operator. In reality, no such copy would be required, we merely include this result as a point of comparison. 
For the queries, the brute force solution performs a linear search checking for each point or element if it intersects the query range.
Evaluating the brute force solution allows us to determine how much of a performance advantage the spatial indices offer and to evaluate whether this advantage is sufficient to justify the memory requirements of the index. 

In each of the result tables below, 
for the build throughput (insertions per second), and the query throughput (queries per second), we classify a library's performance as fast (green cell with bolded text) if is within 10$\times$ of the best performing library for build time or for the given query size. We classify the performance as moderate (yellow with plain text) if it is within 100$\times$. All other results are classified as slow (red with underlined text). For memory usage, we classify any configuration that uses 2$\times$ or less of the raw memory size as having low memory requirements (green with bolded text), any configuration that uses less than 5$\times$ the raw memory size as having moderate memory requirements (yellow with plain text), and all other configurations as having high memory requirements (red with underlined text). In all cases, the raw data size can be seen by looking at the memory requirements for the brute force solution, which makes a copy of the mesh points or elements. 
In the tables, memory (abbreviated mem.) indicates the memory used by the tree, while peak memory indicates the maximum memory used while creating the tree. 
\begin{summary}
    Insight: some libraries temporarily use large amounts of additional memory making them a poor choice for significantly memory constrained environments.
\end{summary}

    \subsubsection{Small Scale Mesh Points}
    
    Each query of extra-small, small, medium, and large size retrieves an average of 0.00104\%, 0.100\%, 1.03\% and 8.85\% of the process's assigned mesh points. Even at this small scale, a few libraries run into errors and are thus not included in the results. First, although FLANN and PCL can use GPUs, with 4 NVIDIA Tesla V100 GPUs with a total of 64 GB RAM, they run into out of memory errors. In addition, since they are the only two libraries that support GPUs it is not possible to perform a true cross-library comparison of this capability. 
    Next, the CGAL range tree experiences out of memory errors. Finally, none of the ANN configurations complete within 24 hours. We therefore include results only for the query sizes that complete in these 24 hours.
    \begin{summary}
        Insight: only two of the libraries can utilize GPUs and they quickly exhaust the GPUs memory. 
        
    \end{summary}
    
    \input{tables/results-small-points}

    The results can be seen in Table~\ref{tab:results-small-points}. A few things should be noted about these results. First, as indicated in the table, the libnabo results are for the tree (heap) with $O(\log n)$ query times rather than the $O(n)$ linear vector heap. 
    Second, \reviewerfix{we use KDTREE2's option to rearrange the data for better performance. We found this offers substantially better query performance with only a slight decrease in build throughput.} Third, nanoflann allows you to create an adaptor for an existing vector so that the storage is not duplicated. We show the results for both no duplicated data (``nanoflann'') and duplicate data (``nanoflann duplicated data''). \new{Finally, three libraries, KDTREE2, libkdtree2, and Octree, have artificially reduced memory requirements because they only support storing coordinate data as floats rather than doubles (and thus offer lower precision).}    

From the results we can see that the CGAL k-d tree has by far the highest insertion throughput, followed by Octree and the PCL octree. However, even one of the slower libraries, \reviewerfix{R-tree with a leaf size of 20, builds in approximately 7 seconds} so it should be kept in mind that (apart from 3DTK and libspatialindex) all of these trees build quite quickly. This highlights the performance benefits of being able to parallelize construction (and querying) across the processes through domain decomposition, rather than attempting to use a single index for the entire mesh (although this does increase the total memory consumption). Shockingly, CGAL is able to create the k-d tree significantly faster than the brute force solution copies the 2D vector of mesh coordinates. There are a few libraries that perform well across all query sizes: ALGLIB, Boost, the CGAL k-d tree, KDTREE2, libkdtree2, Octree, PicoTree and R-tree. 
\begin{summary}
    Insight: query performance is strongly dictated by the number of matching mesh points and when a large number of points are returned, most libraries fail to beat the brute force solution.
\end{summary}

The poor performance on the large queries is a reflection of the fact that all of these libraries are tree based and as such can offer a significant performance advantage when a large portion of the search space can be pruned and a significant performance penalty when large portions of the tree must be traversed. It should also be noted that all of the libraries use far more memory than the theoretical lower bound of linear storage. Almost all of the libraries use between 2-5$\times$ the space needed for just the raw coordinate points (29.2 MB), and several use significantly more than this. 
Only nanoflann with no duplicated storage is able to use anywhere close to linear memory, but it performs more than 10$\times$ slower than the best performing library for all query sizes. 
\footnotetext{A complete explanation of the classification scheme can be found in Section~\ref{sec:performance-evals}.}

    \subsubsection{Small Scale Mesh Elements}
    Each query of extra-small, small, medium, and large size retrieves an average of 0.00213\%, 0.212\%, 1.68\% and 12.68\% of the process's assigned mesh elements. 
    
\input{tables/results-small-bboxes}

The results of this evaluation are shown in Table~\ref{tab:results-small-bboxes}. One library that we test, CGAL's segment tree, runs into out of memory errors, and is therefore not included in the results. 
In these results we again find that a few of the libraries' data structures perform well across the board: Boost, the CGAL R-tree, and the R-tree library. In this case, the libraries overall use (relatively) far less memory, with most using less than 2$\times$ the memory needed for the raw element coordinate data (47.0MB). Unlike with the point results, here all of the best performing libraries outperform brute force at query sizes (including the large queries) since the brute force solution experiences an approximately 2$\times$ slowdown as it now has to consider two points per vector element (the lower and upper corner of the bounding box) rather than just one.

    \input{tables/results-large-points}

    \subsubsection{Large Scale Mesh Points}
    Each query of extra-small, small, medium, and large size retrieves an average of 0.00117\%, 0.0996\%, 1.05\% and 9.84\% of the process's assigned mesh points. Results are shown in Table~\ref{tab:results-large-points}. Although at small scale all of these libraries have good performance for all query sizes, here we see the relative performance decline for the KDTREE2, libkdtree2 and Octree libraries, and for the CGAL k-d tree with the extra-small queries. The rest still perform well across the board. 
    
    \begin{summary}
        Insight: at large scale the libraries outperform the brute force solution at all query scales. 
        These results can help estimate when the brute force solution offers the best performance.
    \end{summary}

    \input{tables/results-large-bboxes}

    \subsubsection{Large Scale Mesh Elements}
     The queries of extra-small, small, medium, and large size retrieve an average of 0.00222\%, 0.120\%, 1.17\% and 10.5\% of the mesh elements assigned to each process. Results are shown in Table~\ref{tab:results-large-bboxes}. Once again, we see that CGAL (this time with an R-tree) struggles with the extra-small queries but has exceptionally fast build throughput, and Boost performs well across the board. Here Boost is the only library that achieves good results in terms of memory requirements given that the raw data takes 472 MB. 
    Once again the R-tree library has significantly slower insertion throughput than the alternative libraries, but offers around a 2$\times$ speedup for the medium and large queries.

\subsection{Scalability Evaluations}
In this section, we present an evaluation of the strong and weak scaling for the libraries.

\subsubsection{Strong Scaling}\label{sec:strong-scaling}
We evaluate the strong scaling by comparing the performance as \reviewerfix{we use 10$\times$ fewer process (16 instead of 160) for the same mesh. As a result,} the data per process increases from small to large (by a factor of 8.28$\times$ for the point tests and 10.0$\times$ for the element tests). The results are shown in Table~\ref{tab:results-strong-scaling-points} and Table~\ref{tab:results-strong-scaling-bboxes}. 
The values shown are the factor difference (the large result divided by the small result). For the performance results, we classify the scaling as good (green with bolded text) if it is better than logarithmic 
$(\frac{1}{\log_{2}8.28}=0.328$ 
for the points and $\frac{1}{\log_{2}10.0}=0.301$ for the elements). We classify the scaling as moderate (yellow with plain text) if it is worse than logarithmic but better than half of logarithmic. All other results are classified as poor (red with underlined text). For the memory results, we classify the result as good (green) if there is a sub-linear increase in memory usage, moderate (yellow) if the increase is greater than linear but less than a 1.5$\times$ increase, and poor (red) if there is a greater than 1.5$\times$ increase.

        \footnotetext{\label{note:performance-classification}A complete explanation of the classification scheme can be found in Section~\ref{sec:performance-evals}.}

\input{tables/results-strong-scaling-points-flipped}

\paragraph{Mesh Points}
As shown in Table~\ref{tab:results-strong-scaling-points}, apart from libkdtree2, the libraries scale very well in terms of insertion throughput and they almost uniformly have a sub-linear increase in memory usage. However, KDTREE2, libkdtree2 and Octree  experience significantly worse than logarithmic scaling for most or all query sizes. Only PicoTree and R-tree experience good scaling across the board, with ALGLIB experiencing good scaling at most query sizes. CGAL's k-d tree experiences moderate scaling for most query sizes.  

\input{tables/results-strong-scaling-bboxes-flipped}

\paragraph{Mesh Elements}
From Table~\ref{tab:results-strong-scaling-bboxes} we can see that all libraries achieve good scaling for insertion throughput and just miss the cutoff for good scaling for the large query size. For the remaining query sizes, all achieve good scaling apart from CGAL's R-tree with the extra-small query size. Interestingly, they all experience a greater than linear increase in memory usage (in contrast to their sub-linear scaling for point storage), and the CGAL R-tree requires almost 20$\times$ the memory despite storing only 10$\times$ as many elements.

\subsubsection{Weak Scaling}\label{sec:weak-scaling}
    To evaluate the weak scaling, we use the same setup used for the performance evaluations with one modification: we only perform 1000 extra-small, small, and medium queries and 100 large queries (1/10th as many queries of each category). This number of queries is sufficient to allow us to compare differences between the libraries. 
    Where more than the baseline number of processes (160 and 16) are used, several processes write and read the same mesh data. This allows us to ensure that the performance differences are the result of the weak scaling changes rather than changes in the mesh data. For space reasons, we only present the results for a representative subset of the libraries. For each of the four experiment setups, we present scaling results for one library that performed (more or less) well across the board and one that performed moderately well. 
    This demonstrates how a library should be expected to scale depending on what performance category it falls into.
    We consider two different weak scaling scenarios: when the number of processes per node is held constant and when the number 
    is varied. \reviewerfix{The results are shown in Tables ~\ref{tab:weak-scaling-fixed-procs-per-node} and ~\ref{tab:weak-scaling-variable-procs-per-node}.} 
    For both scenarios, we compare the results to the performance for the baseline setup used in the performance evaluations (a single node, with 160 processes for a small amount of data per process or 16 processes for a large amount of data). \new{We classify the scaling as good (indicated in green with bold text) if the performance is at most 10\% worse than the baseline, moderate (indicated in yellow with plain text) if it is $(10\%-25\%]$ worse than the baseline, and poor (indicated in red with underlined text) if the performance is over 25\% worse.} In each of the tables, the results indicate the percent change from the baseline with a negative number indicating the operation takes less time than the baseline and a positive number indicating the operation takes longer than the baseline.

\input{tables/weak-scaling-fixed-procs-per-node}

\paragraph{Fixed Number of Processes per Node}
When the number of processes per node is held constant, we would expect to find linear weak scaling (constant performance when the work per process is held constant) since each process performs independent work and does not use shared resources such as the network or file system. To evaluate this, we look at how the performance changes when using 15 or 30 nodes (compared to 1 node). The 15 and 30 node results show very close agreement (within 5\%) so we omit the 15 node results for space reasons. The results are shown in Table~\ref{tab:weak-scaling-fixed-procs-per-node}.

\begin{summary}
Insight: the libraries achieve approximately linear weak scaling in all cases. This indicates our results should hold at any scale if the processes per node and work per process are held constant.
\end{summary}
\footnotetext[3]{\label{note:strong-scaling}A complete explanation of the classification scheme can be found in Section~\ref{sec:strong-scaling}.}

\input{tables/weak-scaling-variable-procs-per-node}

\paragraph{Variable Number of Processes per Node}

Next, we evaluate weak scaling when a different number of processes per node is used. 
When a small amount of data per process is stored, we are already maxing out the processes per node (160). 
We therefore evaluate when 20, 40, 60, and 80 processes per node are used. For space reasons we only present the results for 20 and 80 processes per node.
In this case, we use the same number of total processes (160) but vary the number of nodes. For the case when a large amount of data per process is stored, we only use 16 nodes per process (far below the maximum), and therefore keep the number of nodes constant (one) but increase the number of processes per node. For the mesh point case, we present results for when 32 and 80 processes per node are used. For the mesh elements case we present results for when 32 and 64 processes per node are used because with 80 processes we exhaust the node's memory. The results, shown in Table~\ref{tab:weak-scaling-variable-procs-per-node}, are much more varied. The libraries generally scale well when the number of processes per node is only 2$\times$ different from baseline. However, when the number of processes per node is 4, 5 or 8$\times$ different the results are sometimes greatly affected. This can be seen when using 64 or 80 processes per node with a large amount of data per process and with 20 processes per node with a small amount of data per process. For example, with a small number of points per process, the CGAL k-d tree (for points) and Boost library (for elements) both perform dramatically better when only 20 processes per node are used. With a large amount of data per process, we can see that ALGLIB and Boost both perform substantially worse when the number of processes per node is increased. Thus, some of the libraries are much more sensitive to the number of processes per node that are being utilized.

\footnotetext[4]{\label{note:strong-scaling}A complete explanation of the classification scheme can be found in Section~\ref{sec:strong-scaling}.}

    \footnotetext[5]{\label{note:weak-scaling}A complete explanation of the classification scheme can be found in Section~\ref{sec:weak-scaling}.}.
\vspace{-.3in}

\reviewerfix{
\subsection{Applicability of results}
Although we only use a single mesh type in this evaluation (unstructured, hexahedral), we can use these results to estimate the performance we would get with other mesh types. As shown in Tables~\ref{tab:results-small-points} through \ref{tab:results-large-bboxes}, 
query performance is greatly affected by the query size (percent of stored mesh points or elements that match the query), exhibiting an approximately linear relationship. Therefore, any mesh with more densely clustered points or elements near the range being queried should result in decreased performance. Another factor that should affect performance is the shape of the mesh elements. All of the libraries that support mesh elements (2D or 3D shapes) use axis aligned bounding boxes (AABBs) 
to determine potential intersections. Thus, the more a mesh element deviates from an AABB, the worse the performance because of the increase in tree branches that must be explored and potential matches identified, and the need to use more complex intersection tests to achieve accurate results. 
Beyond these factors, we would not expect much of a performance difference. Although many of the libraries use data structures that are not necessarily balanced (e.g., octrees or k-d trees), almost all of the indices are balanced. This is accomplished through a combination of bulk loading 
and re-balance functions. Therefore, we would not expect to find tree-height related performance differences, or edge cases producing dramatically unbalanced trees. We leave as future work directly evaluating how the use of different mesh types affects these results.}

\begin{summary}
    Insight: our query results can be used to extrapolate expected performance for different mesh types.
\end{summary}

%% file: tables/results-small-points.tex
\begin{table*}[!htbp]
\small
\vspace{-.1in}
\caption{\reviewerfix{Small scale results for mesh points (160 processes, $\approx$1 million points/process). Even at small scale, some libraries perform orders of magnitude better on writing, querying, memory usage. Green is good, yellow is moderate, red is poor\protect\footnotemark.}}
\label{tab:results-small-points}
\begin{tabular}{|l|r|r|r|r|r|r|r|r|}
\hline
\textbf{Configuration} &
  \multicolumn{1}{c|}{\textbf{\begin{tabular}[c]{@{}c@{}}Leaf\\ size\end{tabular}}} &
  \multicolumn{1}{c|}{\textbf{\begin{tabular}[c]{@{}c@{}}Insertions/sec \\ (millions)\end{tabular}}} &
  \multicolumn{1}{c|}{\textbf{\begin{tabular}[c]{@{}c@{}}X-small \\ queries/sec\end{tabular}}} &
  \multicolumn{1}{c|}{\textbf{\begin{tabular}[c]{@{}c@{}}Small \\ queries/sec\end{tabular}}} &
  \multicolumn{1}{c|}{\textbf{\begin{tabular}[c]{@{}c@{}}Medium\\ queries/sec\end{tabular}}} &
  \multicolumn{1}{c|}{\textbf{\begin{tabular}[c]{@{}c@{}}Large\\ queries/sec\end{tabular}}} &
  \multicolumn{1}{c|}{\textbf{\begin{tabular}[c]{@{}c@{}}Mem.\\ (MB)\end{tabular}}} &
  \multicolumn{1}{c|}{\textbf{\begin{tabular}[c]{@{}c@{}}Peak mem. \\ (MB)\end{tabular}}} \\ \hline
\textbf{Brute force} &
  - &
  \cellcolor{my-green}\textbf{5.51} &
  \cellcolor{my-red}{\ul 29.3} &
  \cellcolor{my-red}{\ul 29.3} &
  \cellcolor{my-yellow}29 &
  \cellcolor{my-green}\textbf{27.7} &
  \cellcolor{my-green}\textbf{29.2} &
  \cellcolor{my-green}\textbf{29.2} \\ \hline
 \textbf{3DTK} &
  20 &
  \cellcolor{my-red}{\ul 0.00383} &
  \cellcolor{my-red}{\ul 10.2} &
  \cellcolor{my-red}{\ul 9.32} &
  \cellcolor{my-yellow}7.41 &
  \cellcolor{my-yellow}3.97 &
  \cellcolor{my-red}{\ul 612} &
  \cellcolor{my-red}{\ul 780} \\ \hline
\textbf{ALGLIB} &
  8 &
  \cellcolor{my-yellow}1.11 &
  \cellcolor{my-green}\textbf{36100} &
  \cellcolor{my-green}\textbf{2360} &
  \cellcolor{my-green}\textbf{415} &
  \cellcolor{my-green}\textbf{69.6} &
  \cellcolor{my-red}{\ul 152} &
  \cellcolor{my-red}{\ul 384} \\ \hline
\textbf{ANN k-d tree} &
  200 &
  \cellcolor{my-yellow}1.16 &
  \cellcolor{my-red}{\ul 35.4} &
  \cellcolor{my-red}- &
  \cellcolor{my-red}- &
  \cellcolor{my-red}- &
  \cellcolor{my-green}\textbf{54.6} &
  \cellcolor{my-green}\textbf{54.6} \\ \hline
\textbf{ANN BD-tree} &
  200 &
  \cellcolor{my-yellow}0.704 &
  \cellcolor{my-red}{\ul 36.1} &
  \cellcolor{my-red}- &
  \cellcolor{my-red}- &
  \cellcolor{my-red}- &
  \cellcolor{my-green}\textbf{44.9} &
  \cellcolor{my-green}\textbf{54.6} \\ \hline
\textbf{Boost} &
  200 &
  \cellcolor{my-yellow}2.87 &
  \cellcolor{my-green}\textbf{47900} &
  \cellcolor{my-green}\textbf{3660} &
  \cellcolor{my-green}\textbf{560} &
  \cellcolor{my-green}\textbf{61.1} &
  \cellcolor{my-yellow}79.1 &
  \cellcolor{my-red}{\ul 185} \\ \hline
\textbf{CGAL k-d tree} &
  200 &
  \cellcolor{my-green}\textbf{33.2} &
  \cellcolor{my-green}\textbf{9690} &
  \cellcolor{my-green}\textbf{4060} &
  \cellcolor{my-green}\textbf{625} &
  \cellcolor{my-green}\textbf{62.9} &
  \cellcolor{my-yellow}77.2 &
  \cellcolor{my-yellow}111 \\ \hline
\textbf{FLANN} &
  200 &
  \cellcolor{my-yellow}1.98 &
  \cellcolor{my-red}{\ul 78.6} &
  \cellcolor{my-yellow}55.5 &
  \cellcolor{my-yellow}25.4 &
  \cellcolor{my-yellow}7.64 &
  \cellcolor{my-green}\textbf{54.8} &
  \cellcolor{my-green}\textbf{54.8} \\ \hline
\textbf{KDTREE1} &
  1 &
  \cellcolor{my-yellow}0.388 &
  \cellcolor{my-red}{\ul 99.5} &
  \cellcolor{my-red}{\ul 4.9} &
  \cellcolor{my-red}{\ul 0.901} &
  \cellcolor{my-red}{\ul 0.18} &
  \cellcolor{my-yellow}127 &
  \cellcolor{my-red}{\ul 292} \\ \hline
\textbf{KDTREE2} &
  200 &
  \cellcolor{my-yellow}2.83 &
  \cellcolor{my-green}\textbf{10400} &
  \cellcolor{my-green}\textbf{571} &
  \cellcolor{my-green}\textbf{97.6} &
  \cellcolor{my-green}\textbf{19.2} &
  \cellcolor{my-green}\textbf{45.8} &
  \cellcolor{my-green}\textbf{45.8} \\ \hline
\textbf{KDTREE3} &
  200 &
  \cellcolor{my-yellow}1.65 &
  \cellcolor{my-green}\textbf{5370} &
  \cellcolor{my-yellow}228 &
  \cellcolor{my-yellow}35.5 &
  \cellcolor{my-yellow}6.33 &
  \cellcolor{my-yellow}59.6 &
  \cellcolor{my-yellow}127 \\ \hline
\textbf{KDTREE4} &
  1 &
  \cellcolor{my-red}{\ul 0.297} &
  \cellcolor{my-yellow}563 &
  \cellcolor{my-yellow}65.3 &
  \cellcolor{my-yellow}15.3 &
  \cellcolor{my-yellow}3.55 &
  \cellcolor{my-yellow}115 &
  \cellcolor{my-yellow}115 \\ \hline
\textbf{libkdtree++} &
  1 &
  \cellcolor{my-red}{\ul 0.11} &
  \cellcolor{my-yellow}1120 &
  \cellcolor{my-yellow}95.8 &
  \cellcolor{my-yellow}20.2 &
  \cellcolor{my-yellow}4.49 &
  \cellcolor{my-red}{\ul 272} &
  \cellcolor{my-red}{\ul 272} \\ \hline
\textbf{libkdtree2} &
  1 &
  \cellcolor{my-red}{\ul 0.012} &
  \cellcolor{my-green}\textbf{24000} &
  \cellcolor{my-green}\textbf{994} &
  \cellcolor{my-green}\textbf{109} &
  \cellcolor{my-green}\textbf{12.5} &
  \cellcolor{my-red}{\ul 285} &
  \cellcolor{my-red}{\ul 285} \\ \hline
\textbf{libnabo tree} &
  200 &
  \cellcolor{my-yellow}1.5 &
  \cellcolor{my-red}{\ul 69.1} &
  \cellcolor{my-yellow}64.5 &
  \cellcolor{my-yellow}33 &
  \cellcolor{my-yellow}8.63 &
  \cellcolor{my-yellow}73.4 &
  \cellcolor{my-yellow}94.8 \\ \hline
\textbf{libspatialindex} &
  20 &
  \cellcolor{my-red}{\ul 0.00283} &
  \cellcolor{my-yellow}4290 &
  \cellcolor{my-yellow}330 &
  \cellcolor{my-yellow}40.9 &
  \cellcolor{my-yellow}4.97 &
  \cellcolor{my-yellow}104 &
  \cellcolor{my-red}{\ul 196} \\ \hline
\textbf{nanoflann} &
  200 &
  \cellcolor{my-yellow}0.43 &
  \cellcolor{my-red}{\ul 83.5} &
  \cellcolor{my-yellow}59 &
  \cellcolor{my-yellow}27.4 &
  \cellcolor{my-yellow}8.45 &
  \cellcolor{my-green}\textbf{20.7} &
  \cellcolor{my-green}\textbf{20.7} \\ \hline
\textbf{nanoflann (duplicated data)} &
  200 &
  \cellcolor{my-yellow}1.09 &
  \cellcolor{my-red}{\ul 89} &
  \cellcolor{my-yellow}61.5 &
  \cellcolor{my-yellow}28.3 &
  \cellcolor{my-yellow}8.65 &
  \cellcolor{my-yellow}69.4 &
  \cellcolor{my-yellow}79.1 \\ \hline
\textbf{Octree} &
  200 &
  \cellcolor{my-green}\textbf{4.92} &
  \cellcolor{my-green}\textbf{8180} &
  \cellcolor{my-green}\textbf{526} &
  \cellcolor{my-green}\textbf{98.2} &
  \cellcolor{my-green}\textbf{20.9} &
  \cellcolor{my-green}\textbf{29.5} &
  \cellcolor{my-green}\textbf{29.5} \\ \hline
\textbf{PCL k-d tree} &
  15 &
  \cellcolor{my-yellow}0.708 &
  \cellcolor{my-green}\textbf{5600} &
  \cellcolor{my-yellow}256 &
  \cellcolor{my-yellow}40.4 &
  \cellcolor{my-red}{\ul 0.965} &
  \cellcolor{my-green}\textbf{51.5} &
  \cellcolor{my-yellow}60.5 \\ \hline
\textbf{PCL octree} &
  20 &
  \cellcolor{my-green}\textbf{4.61} &
  \cellcolor{my-red}{\ul 140} &
  \cellcolor{my-yellow}112 &
  \cellcolor{my-green}\textbf{77.1} &
  \cellcolor{my-yellow}4.35 &
  \cellcolor{my-green}\textbf{51.5} &
  \cellcolor{my-yellow}60.5 \\ \hline
\textbf{PicoTree} &
  20 &
  \cellcolor{my-yellow}2.12 &
  \cellcolor{my-green}\textbf{23500} &
  \cellcolor{my-green}\textbf{1620} &
  \cellcolor{my-green}\textbf{311} &
  \cellcolor{my-green}\textbf{57.5} &
  \cellcolor{my-yellow}105 &
  \cellcolor{my-yellow}111 \\ \hline
\textbf{R-tree} &
  20 &
  \cellcolor{my-red}{\ul0.183} &
  \cellcolor{my-green}\textbf{42300} &
  \cellcolor{my-green}\textbf{2720} &
  \cellcolor{my-green}\textbf{447} &
  \cellcolor{my-green}\textbf{68.9} &
  \cellcolor{my-yellow}123 &
  \cellcolor{my-yellow}123 \\ \hline
\textbf{Spatial} &
  1 &
  \cellcolor{my-yellow}0.451 &
  \cellcolor{my-green}\textbf{24000} &
  \cellcolor{my-yellow}330 &
  \cellcolor{my-yellow}32.3 &
  \cellcolor{my-yellow}3.6 &
  \cellcolor{my-yellow}108 &
  \cellcolor{my-red}{\ul 212} \\ \hline
\end{tabular}
\vspace{-.1in}
\end{table*}

%% file: tables/results-small-bboxes.tex
\begin{table*}[!htbp]
\small
\caption{\reviewerfix{Small scale results for hexahedral mesh elements (160 processes, $\approx$1 million elements/process) for the libraries that support 2D or 3D shapes. We find moderately to significantly reduced performance compared to point storage (Table 3), better average performance compared to brute force, and reduced memory overheads. Green is good, yellow is moderate, red is poor\protect\footnotemark.}}
\label{tab:results-small-bboxes}
\begin{tabular}{|l|r|r|r|r|r|r|r|r|}
\hline
\textbf{Configuration} &
  \multicolumn{1}{c|}{\textbf{\begin{tabular}[c]{@{}c@{}}Leaf\\ size\end{tabular}}} &
  \multicolumn{1}{c|}{\textbf{\begin{tabular}[c]{@{}c@{}}Insertions/sec \\ (millions)\end{tabular}}} &
  \multicolumn{1}{c|}{\textbf{\begin{tabular}[c]{@{}c@{}}X-small \\ queries/sec\end{tabular}}} &
  \multicolumn{1}{c|}{\textbf{\begin{tabular}[c]{@{}c@{}}Small \\ queries/sec\end{tabular}}} &
  \multicolumn{1}{c|}{\textbf{\begin{tabular}[c]{@{}c@{}}Medium\\ queries/sec\end{tabular}}} &
  \multicolumn{1}{c|}{\textbf{\begin{tabular}[c]{@{}c@{}}Large\\ queries/sec\end{tabular}}} &
  \multicolumn{1}{c|}{\textbf{\begin{tabular}[c]{@{}c@{}}Mem.\\ (MB)\end{tabular}}} &
  \multicolumn{1}{c|}{\textbf{\begin{tabular}[c]{@{}c@{}}Peak mem. \\ (MB)\end{tabular}}} \\ \hline
\textbf{Brute force} &
  - &
  \cellcolor{my-green}\textbf{2.23} &
  \cellcolor{my-red}{\ul 15.9} &
  \cellcolor{my-red}{\ul 15.8} &
  \cellcolor{my-yellow}15.7 &
  \cellcolor{my-green}\textbf{14.9} &
  \cellcolor{my-green}\textbf{47.0} &
  \cellcolor{my-green}\textbf{47.0} \\ \hline
\textbf{Boost} &
  20 &
  \cellcolor{my-green}\textbf{2.11} &
  \cellcolor{my-green}\textbf{36100} &
  \cellcolor{my-green}\textbf{2250} &
  \cellcolor{my-green}\textbf{270} &
  \cellcolor{my-green}\textbf{33.6} &
  \cellcolor{my-green}\textbf{61.8} &
  \cellcolor{my-yellow}160 \\ \hline
\textbf{CGAL R-tree} &
  1 &
  \cellcolor{my-green}\textbf{15.3} &
  \cellcolor{my-green}\textbf{4170} &
  \cellcolor{my-green}\textbf{1420} &
  \cellcolor{my-green}\textbf{228} &
  \cellcolor{my-green}\textbf{34.8} &
  \cellcolor{my-green}\textbf{75.3} &
  \cellcolor{my-yellow}96.3 \\ \hline
\textbf{libspatialindex} &
  200 &
  \cellcolor{my-red}{\ul 0.051} &
  \cellcolor{my-yellow}2140 &
  \cellcolor{my-yellow}213 &
  \cellcolor{my-yellow}32.8 &
  \cellcolor{my-yellow}4.64 &
  \cellcolor{my-green}\textbf{72.4} &
  \cellcolor{my-yellow}227 \\ \hline
\textbf{R-tree} &
  200 &
  \cellcolor{my-yellow}0.18 &
  \cellcolor{my-green}\textbf{33200} &
  \cellcolor{my-green}\textbf{2460} &
  \cellcolor{my-green}\textbf{419} &
  \cellcolor{my-green}\textbf{67.6} &
  \cellcolor{my-yellow}102 &
  \cellcolor{my-yellow}102 \\ \hline
\textbf{Spatial} &
  1 &
  \cellcolor{my-yellow}0.393 &
  \cellcolor{my-yellow}1010 &
  \cellcolor{my-yellow}233 &
  \cellcolor{my-yellow}40.2 &
  \cellcolor{my-yellow}5.78 &
  \cellcolor{my-yellow}110 &
  \cellcolor{my-yellow}198 \\ \hline
\end{tabular}
\end{table*}

%% file: tables/results-large-points.tex
  \begin{table*}[!htbp]
  \small
  \vspace{-.1in}
  \caption{\reviewerfix{Large scale results for mesh points (16 processes, $\approx$10 million points/process) for libraries that perform best at small scale. Despite the fact that all of these libraries perform well at small scale, significant differences emerge at large scale. Green is good, yellow is moderate, red is poor\textsuperscript{\ref{note:performance-classification}}.}}
  \label{tab:results-large-points}
  \begin{tabular}{|l|r|r|r|r|r|r|r|r|}
  \hline
  \textbf{Configuration} &
    \multicolumn{1}{c|}{\textbf{\begin{tabular}[c]{@{}c@{}}Leaf\\ size\end{tabular}}} &
    \multicolumn{1}{c|}{\textbf{\begin{tabular}[c]{@{}c@{}}Insertions/sec \\ (millions)\end{tabular}}} &
    \multicolumn{1}{c|}{\textbf{\begin{tabular}[c]{@{}c@{}}X-small \\ queries/sec\end{tabular}}} &
    \multicolumn{1}{c|}{\textbf{\begin{tabular}[c]{@{}c@{}}Small \\ queries/sec\end{tabular}}} &
    \multicolumn{1}{c|}{\textbf{\begin{tabular}[c]{@{}c@{}}Medium\\ queries/sec\end{tabular}}} &
    \multicolumn{1}{c|}{\textbf{\begin{tabular}[c]{@{}c@{}}Large\\ queries/sec\end{tabular}}} &
    \multicolumn{1}{c|}{\textbf{\begin{tabular}[c]{@{}c@{}}Mem.\\ (MB)\end{tabular}}} &
    \multicolumn{1}{c|}{\textbf{\begin{tabular}[c]{@{}c@{}}Peak mem. \\ (MB)\end{tabular}}} \\ \hline
  \textbf{Brute force} &
    20 &
    \cellcolor{my-green}\textbf{4.48} &
    \cellcolor{my-red}{\ul 8.08} &
    \cellcolor{my-red}{\ul 8.08} &
    \cellcolor{my-yellow}8.04 &
    \cellcolor{my-green}\textbf{7.71} &
    \cellcolor{my-green}\textbf{240} &
    \cellcolor{my-green}\textbf{240} \\ \hline
  \textbf{ALGLIB} &
    8 &
    \cellcolor{my-yellow}1.03 &
    \cellcolor{my-green}\textbf{29300} &
    \cellcolor{my-green}\textbf{1270} &
    \cellcolor{my-green}\textbf{170} &
    \cellcolor{my-green}\textbf{21.7} &
    \cellcolor{my-yellow}1180 &
    \cellcolor{my-red}{\ul 3170} \\ \hline
  \textbf{Boost} &
    200 &
    \cellcolor{my-yellow}2.44 &
    \cellcolor{my-green}\textbf{19400} &
    \cellcolor{my-green}\textbf{939} &
    \cellcolor{my-green}\textbf{121} &
    \cellcolor{my-green}\textbf{15} &
    \cellcolor{my-yellow}568 &
    \cellcolor{my-red}{\ul 1500} \\ \hline
  \textbf{CGAL k-d tree} &
    200 &
    \cellcolor{my-green}\textbf{29.8} &
    \cellcolor{my-yellow}1660 &
    \cellcolor{my-green}\textbf{1140} &
    \cellcolor{my-green}\textbf{143} &
    \cellcolor{my-green}\textbf{17.6} &
    \cellcolor{my-yellow}537 &
    \cellcolor{my-yellow}886 \\ \hline
  \textbf{\small KDTREE2} &
    200 &
    \cellcolor{my-yellow}2.12 &
    \cellcolor{my-yellow}1610 &
    \cellcolor{my-yellow}63.1 &
    \cellcolor{my-yellow}12.3 &
    \cellcolor{my-yellow}2.65 &
    \cellcolor{my-green}\textbf{374} &
    \cellcolor{my-green}\textbf{374} \\ \hline     
  \textbf{libkdtree2} &
    \multicolumn{1}{r|}{1} &
    \multicolumn{1}{r|}{\cellcolor{my-red}{\ul 0.00285}} &
    \multicolumn{1}{r|}{\cellcolor{my-green}\textbf{7720}} &
    \multicolumn{1}{r|}{\cellcolor{my-green}\textbf{163}} &
    \multicolumn{1}{r|}{\cellcolor{my-yellow}16.1} &
    \multicolumn{1}{r|}{\cellcolor{my-yellow}1.76} &
    \multicolumn{1}{r|}{\cellcolor{my-red}{\ul 2340}} &
    \multicolumn{1}{r|}{\cellcolor{my-red}{\ul 2340}} \\ \hline
  \textbf{Octree} &
    200 &
    \cellcolor{my-green}\textbf{3.28} &
    \cellcolor{my-yellow}1440 &
    \cellcolor{my-yellow}73.1 &
    \cellcolor{my-yellow}14.7 &
    \cellcolor{my-yellow}3.21 &
    \cellcolor{my-green}\textbf{234} &
    \cellcolor{my-green}\textbf{234} \\ \hline
  \textbf{PicoTree} &
    20 &
    \cellcolor{my-yellow}1.53 &
    \cellcolor{my-green}\textbf{16200} &
    \cellcolor{my-green}\textbf{815} &
    \cellcolor{my-green}\textbf{132} &
    \cellcolor{my-green}\textbf{23.5} &
    \cellcolor{my-yellow}852 &
    \cellcolor{my-yellow}886 \\ \hline
  \textbf{R-tree} &
    20 &
    \cellcolor{my-red}{\ul 0.166} &
    \cellcolor{my-green}\textbf{27000} &
    \cellcolor{my-green}\textbf{1290} &
    \cellcolor{my-green}\textbf{175} &
    \cellcolor{my-green}\textbf{22.5} &
    \cellcolor{my-yellow}1020 &
    \cellcolor{my-yellow}1020 \\ \hline
  \end{tabular}
  \vspace{-.05in}
  \end{table*}

%% file: tables/results-large-bboxes.tex
\begin{table*}[!htbp]
\small
\caption{\reviewerfix{Large scale results for hexahedral mesh elements (16 processes, $\approx$11 million elements/process) for libraries that perform best at small scale. CGAL offers the best write performance, R-tree generally offers the best read performance, and Boost does well across the board. Green is good, yellow is moderate, red is poor\protect\footnoteref{note:performance-classification}.}}
\label{tab:results-large-bboxes}
\begin{tabular}{|l|r|r|r|r|r|r|r|r|}
\hline
\textbf{Configuration} &
  \multicolumn{1}{c|}{\textbf{\begin{tabular}[c]{@{}c@{}}Leaf\\ size\end{tabular}}} &
  \multicolumn{1}{c|}{\textbf{\begin{tabular}[c]{@{}c@{}}Insertions/sec \\ (millions)\end{tabular}}} &
  \multicolumn{1}{c|}{\textbf{\begin{tabular}[c]{@{}c@{}}X-small \\ queries/sec\end{tabular}}} &
  \multicolumn{1}{c|}{\textbf{\begin{tabular}[c]{@{}c@{}}Small \\ queries/sec\end{tabular}}} &
  \multicolumn{1}{c|}{\textbf{\begin{tabular}[c]{@{}c@{}}Medium\\ queries/sec\end{tabular}}} &
  \multicolumn{1}{c|}{\textbf{\begin{tabular}[c]{@{}c@{}}Large\\ queries/sec\end{tabular}}} &
  \multicolumn{1}{c|}{\textbf{\begin{tabular}[c]{@{}c@{}}Mem.\\ (MB)\end{tabular}}} &
  \multicolumn{1}{c|}{\textbf{\begin{tabular}[c]{@{}c@{}}Peak mem. \\ (MB)\end{tabular}}} \\ \hline
 \textbf{Brute Force} &
  {-} &
  \cellcolor{my-green}\textbf{2.43} &
  \cellcolor{my-red}{\ul 4.34} &
  \cellcolor{my-red}{\ul 4.34} &
  \cellcolor{my-yellow}4.3 &
  \cellcolor{my-green}\textbf{3.97} &
  \cellcolor{my-green}\textbf{473} &
  \cellcolor{my-green}\textbf{473} \\ \hline
\textbf{Boost} &
  20 &
  \cellcolor{my-green}\textbf{1.78} &
  \cellcolor{my-green}\textbf{21200} &
  \cellcolor{my-green}\textbf{773} &
  \cellcolor{my-green}\textbf{80.7} &
  \cellcolor{my-green}\textbf{8.95} &
  \cellcolor{my-green}\textbf{697} &
  \cellcolor{my-yellow}1950 \\ \hline
\textbf{CGAL R-tree} &
  1 &
  \cellcolor{my-green}\textbf{15.3} &
  \cellcolor{my-yellow}424 &
  \cellcolor{my-green}\textbf{530} &
  \cellcolor{my-green}\textbf{70.5} &
  \cellcolor{my-green}\textbf{8.87} &
  \cellcolor{my-yellow}1490 &
  \cellcolor{my-yellow}1490 \\ \hline
\textbf{R-tree} &
  200 &
  \cellcolor{my-yellow}0.163 &
  \cellcolor{my-green}\textbf{17400} &
  \cellcolor{my-green}\textbf{968} &
  \cellcolor{my-green}\textbf{142} &
  \cellcolor{my-green}\textbf{19.2} &
  \cellcolor{my-yellow}1060 &
  \cellcolor{my-yellow}1060 \\ \hline
\end{tabular}
\vspace{-.1in}
\end{table*}

%% file: tables/results-strong-scaling-points-flipped.tex
\begin{table*}[!htbp]
\small
\vspace{-.05in}
\caption{\reviewerfix{Strong scaling results for mesh points. Few libraries have good query scalability. Green is good, yellow is moderate, red is poor\protect\footnotemark.}}
\label{tab:results-strong-scaling-points}
\begin{tabular}{|l|r|r|r|r|r|r|r|r|}
\hline
\textbf{Configuration} &
  \multicolumn{1}{c|}{\textbf{\begin{tabular}[c]{@{}c@{}}Leaf\\ size\end{tabular}}} &
  \multicolumn{1}{c|}{\textbf{Insertions/sec}} &
  \multicolumn{1}{c|}{\textbf{\begin{tabular}[c]{@{}c@{}}X-small\\ queries/sec\end{tabular}}} &
  \multicolumn{1}{c|}{\textbf{\begin{tabular}[c]{@{}c@{}}Small \\ queries/sec\end{tabular}}} &
  \multicolumn{1}{c|}{\textbf{\begin{tabular}[c]{@{}c@{}}Medium\\ queries/sec\end{tabular}}} &
  \multicolumn{1}{c|}{\textbf{\begin{tabular}[c]{@{}c@{}}Large\\ queries/sec\end{tabular}}} &
  \multicolumn{1}{c|}{\textbf{\begin{tabular}[c]{@{}c@{}}Mem.\\ (MB)\end{tabular}}} &
  \multicolumn{1}{c|}{\textbf{\begin{tabular}[c]{@{}c@{}}Peak mem. \\ (MB)\end{tabular}}} \\ \hline
\textbf{Brute force} &
  - &
  \cellcolor{my-green}\textbf{0.81} &
  \cellcolor{my-yellow}0.28 &
  \cellcolor{my-yellow}0.28 &
  \cellcolor{my-yellow}0.28 &
  \cellcolor{my-yellow}0.28 &
  \cellcolor{my-green}\textbf{8.22} &
  \cellcolor{my-green}\textbf{8.22} \\ \hline
\textbf{ALGLIB} &
  8 &
  \cellcolor{my-green}\textbf{0.93} &
  \cellcolor{my-green}\textbf{0.81} &
  \cellcolor{my-green}\textbf{0.54} &
  \cellcolor{my-green}\textbf{0.41} &
  \cellcolor{my-yellow}0.31 &
  \cellcolor{my-green}\textbf{7.76} &
  \cellcolor{my-green}\textbf{8.26} \\ \hline
\textbf{Boost} &
  200 &
  \cellcolor{my-green}\textbf{0.85} &
  \cellcolor{my-green}\textbf{0.41} &
  \cellcolor{my-yellow}0.26 &
  \cellcolor{my-yellow}0.22 &
  \cellcolor{my-yellow}0.25 &
  \cellcolor{my-green}\textbf{7.18} &
  \cellcolor{my-green}\textbf{8.11} \\ \hline
\textbf{CGAL k-d tree} &
  200 &
  \cellcolor{my-green}\textbf{0.90} &
  \cellcolor{my-yellow}0.17 &
  \cellcolor{my-yellow}0.28 &
  \cellcolor{my-yellow}0.23 &
  \cellcolor{my-yellow}0.28 &
  \cellcolor{my-green}\textbf{6.96} &
  \cellcolor{my-green}\textbf{7.98} \\ \hline
\textbf{\small KDTREE2} &
  200 &
  \cellcolor{my-green}\textbf{0.75} &
  \cellcolor{my-red}{\ul 0.15} &
  \cellcolor{my-red}{\ul 0.11} &
  \cellcolor{my-red}{\ul 0.13} &
  \cellcolor{my-red}{\ul 0.14} &
  \cellcolor{my-green}\textbf{8.16} &
  \cellcolor{my-green}\textbf{8.16} \\ \hline
\textbf{libkdtree2} &
  1 &
  \cellcolor{my-yellow}0.24 &
  \cellcolor{my-yellow}0.32 &
  \cellcolor{my-red}{\ul 0.16} &
  \cellcolor{my-red}{\ul 0.15} &
  \cellcolor{my-red}{\ul 0.14} &
  \cellcolor{my-green}\textbf{8.21} &
  \cellcolor{my-green}\textbf{8.21} \\ \hline
\textbf{Octree} &
  200 &
  \cellcolor{my-green}\textbf{0.67} &
  \cellcolor{my-yellow}0.18 &
  \cellcolor{my-red}{\ul 0.14} &
  \cellcolor{my-red}{\ul 0.15} &
  \cellcolor{my-red}{\ul 0.15} &
  \cellcolor{my-green}\textbf{7.95} &
  \cellcolor{my-green}\textbf{7.95} \\ \hline
\textbf{PicoTree} &
  20 &
  \cellcolor{my-green}\textbf{0.72} &
  \cellcolor{my-green}\textbf{0.69} &
  \cellcolor{my-green}\textbf{0.50} &
  \cellcolor{my-green}\textbf{0.42} &
  \cellcolor{my-green}\textbf{0.41} &
  \cellcolor{my-green}\textbf{8.11} &
  \cellcolor{my-green}\textbf{7.98} \\ \hline
\textbf{R-tree} &
  20 &
  \cellcolor{my-green}\textbf{0.91} &
  \cellcolor{my-green}\textbf{0.64} &
  \cellcolor{my-green}\textbf{0.47} &
  \cellcolor{my-green}\textbf{0.39} &
  \cellcolor{my-green}\textbf{0.33} &
  \cellcolor{my-green}\textbf{8.29} &
  \cellcolor{my-green}\textbf{8.29} \\ \hline
\end{tabular}
\vspace{-.1in}
\end{table*}

%% file: tables/results-strong-scaling-bboxes-flipped.tex
\begin{table*}[!htbp]
\small
\caption{\reviewerfix{Strong scaling results for hexahedral mesh elements. The libraries generally have good scalability apart from large queries and memory overheads. Green is good, yellow is moderate, red is poor\protect\footnotemark}.}
\label{tab:results-strong-scaling-bboxes}
\begin{tabular}{|l|r|r|r|r|r|r|r|r|}
\hline
\textbf{Configuration} &
  \multicolumn{1}{c|}{\textbf{\begin{tabular}[c]{@{}c@{}}Leaf\\ size\end{tabular}}} &
  \multicolumn{1}{c|}{\textbf{Insertions/sec}} &
  \multicolumn{1}{c|}{\textbf{\begin{tabular}[c]{@{}c@{}}X-small \\ queries/sec\end{tabular}}} &
  \multicolumn{1}{c|}{\textbf{\begin{tabular}[c]{@{}c@{}}Small \\ queries/sec\end{tabular}}} &
  \multicolumn{1}{c|}{\textbf{\begin{tabular}[c]{@{}c@{}}Medium\\ queries/sec\end{tabular}}} &
  \multicolumn{1}{c|}{\textbf{\begin{tabular}[c]{@{}c@{}}Large\\ queries/sec\end{tabular}}} &
  \multicolumn{1}{c|}{\textbf{\begin{tabular}[c]{@{}c@{}}Mem.\\ (MB)\end{tabular}}} &
  \multicolumn{1}{c|}{\textbf{\begin{tabular}[c]{@{}c@{}}Peak mem. \\ (MB)\end{tabular}}} \\ \hline
\textbf{Brute force} &
  - &
  \cellcolor{my-green}\textbf{1.09} &
  \cellcolor{my-yellow}0.27 &
  \cellcolor{my-yellow}0.27 &
  \cellcolor{my-yellow}0.27 &
  \cellcolor{my-yellow}0.27 &
  \cellcolor{my-yellow}10.0 &
  \cellcolor{my-yellow}10.0 \\ \hline
\textbf{Boost} &
  20 &
  \cellcolor{my-green}\textbf{0.84} &
  \cellcolor{my-green}\textbf{0.59} &
  \cellcolor{my-green}\textbf{0.34} &
  \cellcolor{my-green}\textbf{0.30} &
  \cellcolor{my-yellow}0.27 &
  \cellcolor{my-yellow}11.3 &
  \cellcolor{my-yellow}12.2 \\ \hline
\textbf{CGAL R-tree} &
  20 &
  \cellcolor{my-green}\textbf{1.00} &
  \cellcolor{my-red}{\ul 0.10} &
  \cellcolor{my-green}\textbf{0.37} &
  \cellcolor{my-green}\textbf{0.31} &
  \cellcolor{my-yellow}0.25 &
  \cellcolor{my-red}{\ul 19.8} &
  \cellcolor{my-red}{\ul 15.5} \\ \hline
\textbf{R-tree} &
  200 &
  \cellcolor{my-green}\textbf{0.91} &
  \cellcolor{my-green}\textbf{0.52} &
  \cellcolor{my-green}\textbf{0.39} &
  \cellcolor{my-green}\textbf{0.34} &
  \cellcolor{my-yellow}0.28 &
  \cellcolor{my-yellow}10.4 &
  \cellcolor{my-yellow}10.4 \\ \hline
\end{tabular}
\end{table*}

%% file: tables/weak-scaling-fixed-procs-per-node.tex
\begin{table*}[!htbp]
\small
\vspace{-.1in}
    \caption{\reviewerfix{Weak scaling results for a representative subset of the libraries using a fixed number of processes per node. As expected, the libraries all exhibit approximately linear weak scaling. Green is good, yellow is moderate\protect\footnotemark[5].}}
\label{tab:weak-scaling-fixed-procs-per-node}
\begin{tabular}{|l|r|l|c|r|r|r|r|r|r|r|r|}
\hline
\multicolumn{1}{|c|}{\textbf{Configuration}} &
  \multicolumn{1}{c|}{\textbf{\begin{tabular}[c]{@{}c@{}}Leaf \\ Size\end{tabular}}} &
  \multicolumn{1}{c|}{\textbf{Data}} &
  \multicolumn{1}{c|}{\textbf{\begin{tabular}[c]{@{}c@{}}Data per\\ proc\end{tabular}}} &
  \multicolumn{1}{c|}{\textbf{Procs}} &
  \textbf{Nodes} &
  \multicolumn{1}{c|}{\textbf{\begin{tabular}[c]{@{}c@{}}Procs per\\ node\end{tabular}}} &
  \multicolumn{1}{c|}{\textbf{Insertions}} &
  \multicolumn{1}{c|}{\textbf{\begin{tabular}[c]{@{}c@{}}X-small\\ queries\end{tabular}}} &
  \multicolumn{1}{c|}{\textbf{\begin{tabular}[c]{@{}c@{}}Small \\ queries\end{tabular}}} &
  \multicolumn{1}{c|}{\textbf{\begin{tabular}[c]{@{}c@{}}Medium \\ queries\end{tabular}}} &
  \multicolumn{1}{c|}{\textbf{\begin{tabular}[c]{@{}c@{}}Large\\ queries\end{tabular}}} \\ \hline
\textbf{CGAL k-d tree} &
  20 &
  Points &
  Small &
  4800 &
  30 &
  160 &
  \cellcolor{my-green}\textbf{4.7\%} &
  \cellcolor{my-green}\textbf{-6.4\%} &
  \cellcolor{my-green}\textbf{-1.1\%} &
  \cellcolor{my-green}\textbf{2.0\%} &
  \cellcolor{my-green}\textbf{2.2\%} \\ \hline
\textbf{KDTREE} &
  1 &
  Points &
  Small &
  4800 &
  30 &
  160 &
  \cellcolor{my-green}\textbf{0.4\%} &
  \cellcolor{my-green}\textbf{-4.6\%} &
  \cellcolor{my-green}\textbf{0.7\%} &
  \cellcolor{my-green}\textbf{-0.6\%} &
  \cellcolor{my-green}\textbf{-2.0\%} \\ \hline
\textbf{Boost} &
  200 &
  Elems &
  Small &
  4800 &
  30 &
  160 &
  \cellcolor{my-green}\textbf{-11.0\%} &
  \cellcolor{my-green}\textbf{-2.9\%} &
  \cellcolor{my-green}\textbf{4.8\%} &
  \cellcolor{my-green}\textbf{-2.7\%} &
  \cellcolor{my-green}\textbf{-7.9\%} \\ \hline
\textbf{Spatial} &
  20 &
  Elems &
  Small &
  4800 &
  30 &
  160 &
  \cellcolor{my-green}\textbf{-2.1\%} &
  \cellcolor{my-green}\textbf{-48.2\%} &
  \cellcolor{my-green}\textbf{-2.6\%} &
  \cellcolor{my-green}\textbf{-0.1\%} &
  \cellcolor{my-green}\textbf{-2.2\%} \\ \hline
\textbf{ALGLIB} &
  8 &
  Points &
  Large &
  480 &
  30 &
  16 &
  \cellcolor{my-green}\textbf{0.5\%} &
  \cellcolor{my-green}\textbf{7.9\%} &
  \cellcolor{my-green}\textbf{3.4\%} &
  \cellcolor{my-green}\textbf{5.5\%} &
  \cellcolor{my-green}\textbf{0.3\%} \\ \hline
\textbf{KDTREE2} &
  200 &
  Points &
  Large &
  480 &
  30 &
  16 &
  \cellcolor{my-green}\textbf{0.0\%} &
  \cellcolor{my-green}\textbf{1.4\%} &
  \cellcolor{my-green}\textbf{2.7\%} &
  \cellcolor{my-green}\textbf{2.7\%} &
  \cellcolor{my-green}\textbf{2.3\%} \\ \hline
\textbf{Boost} &
  200 &
  Elems &
  Large &
  480 &
  30 &
  16 &
      \cellcolor{my-yellow}11.6\% &
      \cellcolor{my-yellow}19.4\% &
      \cellcolor{my-green}\textbf{9.8\%} &
      \cellcolor{my-green}\textbf{8.7\%} &
      \cellcolor{my-green}\textbf{1.5\%} \\ \hline
\textbf{R-tree} &
  200 &
  Elems &
  Large &
  480 &
  30 &
  16 &
  \cellcolor{my-green}\textbf{4.7\%} &
  \cellcolor{my-green}\textbf{0.9\%} &
  \cellcolor{my-green}\textbf{-6.3\%} &
  \cellcolor{my-green}\textbf{-1.7\%} &
  \cellcolor{my-green}\textbf{-5.5\%} \\ \hline
\end{tabular}
\vspace{-.1in}
\end{table*}

%% file: tables/weak-scaling-variable-procs-per-node.tex
\begin{table*}[!htbp]
\small
\caption{\reviewerfix{Weak scaling results for a representative subset of the libraries using a variable number of processes per node . Some libraries are more sensitive than others to the number of processes per node. Green is good, yellow is moderate, red is poor\protect\footnoteref{note:weak-scaling}.}}
\label{tab:weak-scaling-variable-procs-per-node}
\begin{tabular}{|l|r|c|c|r|r|r|r|r|r|r|r|}
\hline
\multicolumn{1}{|c|}{\textbf{Configuration}} &
  \multicolumn{1}{c|}{\textbf{\begin{tabular}[c]{@{}c@{}}Leaf\\ Size\end{tabular}}} &
  \textbf{Data} &
  \textbf{\begin{tabular}[c]{@{}c@{}}Data per\\ proc\end{tabular}} &
  \textbf{Nodes} &
  \multicolumn{1}{c|}{\textbf{Procs}} &
  \textbf{\begin{tabular}[c]{@{}r@{}}Procs per\\ node\end{tabular}} &
  \multicolumn{1}{c|}{\textbf{Insertions}} &
  \multicolumn{1}{c|}{\textbf{\begin{tabular}[c]{@{}c@{}}X-small\\ queries\end{tabular}}} &
  \multicolumn{1}{c|}{\textbf{\begin{tabular}[c]{@{}c@{}}Small \\ queries\end{tabular}}} &
  \multicolumn{1}{c|}{\textbf{\begin{tabular}[c]{@{}c@{}}Medium \\ queries\end{tabular}}} &
  \multicolumn{1}{c|}{\textbf{\begin{tabular}[c]{@{}c@{}}Large\\ queries\end{tabular}}} \\ \hline
\textbf{CGAL k-d tree} &
  20 &
  Points &
  Small &
  2 &
  160 &
  80 &
  \cellcolor{my-yellow}21.7\% &
  \cellcolor{my-green}\textbf{1.3\%} &
  \cellcolor{my-green}\textbf{6.6\%} &
  \cellcolor{my-green}\textbf{3.8\%} &
  \cellcolor{my-green}\textbf{-16.4\%} \\ \hline
\textbf{CGAL k-d tree} &
  20 &
  Points &
  Small &
  8 &
  160 &
  20 &
  \cellcolor{my-red}{\ul 38.9\%} &
  \cellcolor{my-green}\textbf{-48.5\%} &
  \cellcolor{my-green}\textbf{-19.7\%} &
  \cellcolor{my-green}\textbf{-28.5\%} &
  \cellcolor{my-green}\textbf{-79.2\%} \\ \hline
\textbf{KDTREE} &
  1 &
  Points &
  Small &
  2 &
  160 &
  80 &
  \cellcolor{my-green}\textbf{7.7\%} &
  \cellcolor{my-green}\textbf{0.4\%} &
  \cellcolor{my-green}\textbf{1.6\%} &
  \cellcolor{my-green}\textbf{-0.3\%} &
  \cellcolor{my-green}\textbf{-1.9\%} \\ \hline
\textbf{KDTREE} &
  1 &
  Points &
  Small &
  8 &
  160 &
  20 &
  \cellcolor{my-green}\textbf{4.7\%} &
  \cellcolor{my-green}\textbf{-5.9\%} &
  \cellcolor{my-green}\textbf{-4.5\%} &
  \cellcolor{my-green}\textbf{-5.7\%} &
  \cellcolor{my-green}\textbf{-6.7\%} \\ \hline
\textbf{Boost} &
  200 &
  Elems &
  Small &
  2 &
  160 &
  80 &
  \cellcolor{my-green}\textbf{9.1\%} &
  \cellcolor{my-yellow}15.4\% &
  \cellcolor{my-red}{\ul 33.7\%} &
  \cellcolor{my-yellow}19.6\% &
  \cellcolor{my-yellow}11.6\% \\ \hline
\textbf{Boost} &
  200 &
  Elems &
  Small &
  8 &
  160 &
  20 &
  \cellcolor{my-green}\textbf{-17.5\%} &
  \cellcolor{my-green}\textbf{-55.1\%} &
  \cellcolor{my-green}\textbf{-49.4\%} &
  \cellcolor{my-green}\textbf{-67.0\%} &
  \cellcolor{my-green}\textbf{-43.5\%} \\ \hline
\textbf{Spatial} &
  20 &
  Elems &
  Small &
  2 &
  160 &
  80 &
  \cellcolor{my-yellow}12.5\% &
  \cellcolor{my-green}\textbf{1.0\%} &
  \cellcolor{my-green}\textbf{0.9\%} &
  \cellcolor{my-green}\textbf{-0.7\%} &
  \cellcolor{my-green}\textbf{-2.7\%} \\ \hline
\textbf{Spatial} &
  20 &
  Elems &
  Small &
  8 &
  160 &
  20 &
  \cellcolor{my-green}\textbf{-12.2\%} &
  \cellcolor{my-green}\textbf{-56.9\%} &
  \cellcolor{my-green}\textbf{-19.6\%} &
  \cellcolor{my-green}\textbf{-18.0\%} &
  \cellcolor{my-green}\textbf{-20.5\%} \\ \hline
\textbf{ALGLIB} &
  8 &
  Points &
  Large &
  1 &
  32 &
  32 &
  \cellcolor{my-green}\textbf{-0.9\%} &
  \cellcolor{my-green}\textbf{-1.9\%} &
  \cellcolor{my-green}\textbf{1.3\%} &
  \cellcolor{my-green}\textbf{-4.4\%} &
  \cellcolor{my-green}\textbf{-14.7\%} \\ \hline
\textbf{ALGLIB} &
  8 &
  Points &
  Large &
  1 &
  80 &
  80 &
  \cellcolor{my-yellow}17.9\% &
  \cellcolor{my-red}{\ul 26.3\%} &
  \cellcolor{my-yellow}22.2\% &
  \cellcolor{my-yellow}16.0\% &
  \cellcolor{my-green}\textbf{2.5\%} \\ \hline
\textbf{KDTREE2} &
  200 &
  Points &
  Large &
  1 &
  32 &
  32 &
  \cellcolor{my-green}\textbf{-2.7\%} &
  \cellcolor{my-green}\textbf{-2.4\%} &
  \cellcolor{my-green}\textbf{0.1\%} &
  \cellcolor{my-green}\textbf{-0.2\%} &
  \cellcolor{my-green}\textbf{-6.8\%} \\ \hline
\textbf{KDTREE2} &
  200 &
  Points &
  Large &
  1 &
  80 &
  80 &
  \cellcolor{my-green}\textbf{4.9\%} &
  \cellcolor{my-green}\textbf{0.8\%} &
  \cellcolor{my-green}\textbf{6.2\%} &
  \cellcolor{my-green}\textbf{6.2\%} &
  \cellcolor{my-green}\textbf{-2.5\%} \\ \hline
\textbf{Boost} &
  200 &
  Elems &
  Large &
  1 &
  32 &
  32 &
  \cellcolor{my-yellow}14.7\% &
  \cellcolor{my-yellow}20.3\% &
  \cellcolor{my-green}\textbf{6.1\%} &
  \cellcolor{my-green}\textbf{-0.5\%} &
  \cellcolor{my-green}\textbf{-6.5\%} \\ \hline
\textbf{Boost} &
  200 &
  Elems &
  Large &
  1 &
  64 &
  64 &
  \cellcolor{my-red}{\ul 25.8\%} &
  \cellcolor{my-red}{\ul 40.7\%} &
  \cellcolor{my-yellow}24.0\% &
  \cellcolor{my-yellow}14.1\% &
  \cellcolor{my-green}\textbf{-0.9\%} \\ \hline
\textbf{R-tree} &
  200 &
  Elems &
  Large &
  1 &
  32 &
  32 &
  \cellcolor{my-green}\textbf{7.1\%} &
  \cellcolor{my-green}\textbf{-7.7\%} &
  \cellcolor{my-green}\textbf{-18.4\%} &
  \cellcolor{my-green}\textbf{-13.4\%} &
  \cellcolor{my-green}\textbf{-19.5\%} \\ \hline
\textbf{R-tree} &
  200 &
  Elems &
  Large &
  1 &
  64 &
  64 &
  \cellcolor{my-green}\textbf{7.2\%} &
  \cellcolor{my-yellow}10.8\% &
  \cellcolor{my-green}\textbf{5.8\%} &
  \cellcolor{my-yellow}12.6\% &
  \cellcolor{my-green}\textbf{5.4\%} \\ \hline
\end{tabular}
\vspace{-.1in}
\end{table*}

%% file: discussion-future-work.tex
\section{Discussion, Insights and Future Work}
\label{sec:future-work}
In this work, we set out to answer the following questions regarding the 20 free, open-source C/C++ libraries that support range queries:
    \begin{enumerate}
        \item Which of the implementations are viable in an HPC setting?
        \item How do these libraries compare in terms of build time, query time, and memory usage at different scales? 
        \item What are other factors in deciding which library to use?
        \item Is there a single overall best solution?
        \item When does a brute force solution offer the best performance?
    \end{enumerate}

The second question has been thoroughly addressed in the evaluation section (see Section~\ref{sec:eval}). In this section we summarize answers to the remaining questions using the evaluation results \reviewerfix{and offer a number of insights that can assist both HPC application scientists and spatial index developers.} We also present areas for future work.

\subsection{Which of the Implementations Are Viable in an HPC Setting?}
\new{As mentioned above, to be viable in an HPC setting, a spatial indexing library must have fast performance for building the index and performing queries, good scalability, and moderate memory usage. In this section, we discuss which libraries best meet these criteria.}

Overall, the results show that for point storage, ALGLIB, Boost, CGAL's k-d tree, PicoTree and R-tree achieve the best query results. 
KDTREE2, libkdtree2 and Octree perform well at small scale but achieve worse performance at large scale. Of these five best performing libraries, R-tree experiences orders of magnitude worse insertion throughput. Therefore, if the application will be performing a small number of queries (thus resulting in the index construction being a larger proportion of work), the R-tree library should not be used, and CGAL's k-dtree, which offers almost an order of magnitude improvement over the next best library, should be strongly considered. In addition, both the R-tree library and ALGLIB require around 4$\times$ more memory than is required for the raw data points. Therefore, these are not viable solutions for particularly memory constrained environments. 
However, for point storage the R-tree library does offer some of the fastest query throughputs. In fact, if many medium and large queries will be performed at large scale the R-tree library offers the fastest solution providing nearly double the query throughput for medium and large queries.

For storing mesh elements, Boost, CGAL's R-tree and the R-tree library offer the best performance. CGAL's R-tree offers close to a 10$\times$ improvement in insertion throughput over Boost and a 100$\times$ throughput improvement over R-tree both at large and small scales, but generally offers the worst performance of the three for all query sizes (and particularly poor performance for extra-small queries). It is therefore a viable solution if a relatively small number of queries will be performed. Boost has by far the smallest memory usage (although it has the highest peak memory usage) and is thus the best solution for moderately memory constrained environments that can tolerate temporary memory pressure. Finally, the R-tree library almost always offers the best query performance for any query size, performing around 2$\times$ better than both Boost and CGAL for the medium and large queries at both small scale and large scale. It is therefore a viable solution when slightly higher memory usage can be tolerated, and a large number of queries will be performed. 

It is worth emphasizing that 
all of the best performing libraries use significantly more memory than the raw data size and therefore are not suitable for severely memory constrained environments. If memory is severely constrained, we would recommend nanoflann with a data adapter (which is very simple to write), which results in sub-linear memory usage, or the Octree library for point storage and the Boost library for element storage. 
For severely space constrained environments, there is another simple option: store floats rather than doubles. This will cut memory requirements approximately in half. We expect that four bytes worth of precision will be sufficient for range queries in most cases given that the spatial index is not taking the place of the mesh coordinate storage, but rather is offering an additional data structure on top (users therefore are not losing mesh precision, simply a small amount of query precision). 
It should also be kept in mind that it is the number of mesh points per process that determines the total structure size rather than the dataset size. Most simulations store a large number of variables (anywhere from 5-100) and a large number of timesteps (anywhere from 100-100,000) and the number of mesh coordinates or elements will therefore be a relatively small fraction of the overall dataset.
It is also worth emphasizing that, given how quickly these structures can be created, they can easily be generated on the fly and therefore will not require any long-term storage.

\subsection{What Are Other Considerations in Deciding Which Library to Use?} 
There are a large number of different factors to consider when deciding which library to use.
Due to space constraints, we will only discuss a few of the most important additional factors here, but a more complete overview can be found in the Git repository for this project~\cite{lawson:2021:range_tree_benchmarking_suite_git_repo}.

\subsubsection{Ease of Use}
All of the libraries are easy to use when compiling a project. CGAL, PicoTree and R-tree are header only libraries. For ALGLIB you simply link your executable to the necessary sources, and for Boost you link to the library, which is widely available and easy to compile. However, the libraries differ greatly in terms of their flexibility for data storage. ALGLIB can only store numeric data (coordinates, integer tags, and N additional real values per coordinate). 
Boost and CGAL offer significant flexibility in what data can be stored, but require the data to be based on classes defined in the library. For Boost this requires containing a boost::geometry::model::point/box as part of what is inserted (for point and box storage). For CGAL, this is more complex. 
Here is an overview of the definitions we used to create a k-d tree that contains 3D points 
and query it using a bounding box.
\begin{lstlisting}[language=C++]
typedef CGAL::Simple_cartesian<double> cgal_kernel;
typedef cgal_kernel::Point_3 cgal_pt;
typedef boost::tuple<cgal_pt,uint32_t> cgal_pt_w_index;
typedef CGAL::Search_traits_3<cgal_kernel> cgal_traits;
typedef CGAL::Search_traits_adapter<cgal_pt_w_index,
    CGAL::Nth_of_tuple_property_map<0, cgal_p_w_index>, 
    cgal_traits> Traits;
typedef CGAL::Sliding_midpoint<Traits> cgal_split_rule;
typedef CGAL::Kd_tree<Traits> cgal_kd_tree;
typedef CGAL::Fuzzy_iso_box<Traits> cgal_box;
\end{lstlisting}
CGAL is nearly infinitely flexible, but the extensive documentation only demonstrates a small fraction of how its capabilities work and can be combined, resulting in a substantial challenge even for experienced coders. 
PicoTree allows you to very easily create a custom class and is incredibly simple to use overall. The R-tree library is the simplest to adapt to the evaluated use-case since it allows you to indicate 
an element data type, and thus does not require any modifications (or custom classes) for use in this work.

\subsubsection{Support and Documentation}
Of the five best performing libraries, only ALGLIB offers commercial support. ALGLIB's commercial version also provides multithreading support and SIMD usage. ALGLIB, Boost and CGAL all offer substantial documentation. PicoTree and R-tree offer little to no documentation but this is somewhat offset by the fact that they have much smaller and simpler code bases, making it much easier to find relevant functions.

\subsubsection{Static vs. Dynamic Trees}
ALGLIB offers only static trees, Boost offers dynamic trees (with support for bulk loading), CGAL's k-d tree supports deletion but is not self balancing, CGAL's R-tree is static, PicoTree offers static trees only, and R-tree offers dynamic trees. Therefore, only Boost, CGAL's k-d tree or R-tree would be suitable for dynamic environments. 

\new{
\subsubsection{Theoretical Performance Bounds}
All of the five best performing libraries for storing points use either a k-d tree (ALGLIB, CGAL's k-d tree, PicoTree) or an R-tree (Boost, the R-tree library). For element storage, all of the best performing libraries use an R-tree (Boost, CGAL's R-tree and the R-tree library). Both k-d trees and R-trees can answer a range query in $O(n^{1-1/d} + k)$ time~\cite{agarwal:1996:range-searching} whereas octrees can answer range queries in $O(\log n)$ time~\cite{jackins:1980:octree-first}. Therefore, it is perhaps a bit surprising that none of the best performing implementations are octrees. Given the wide range of performance we see across the libraries for each tree type, we should expect the library's implementation to be a much greater factor in determining performance than the theoretical bounds of the data structure.
}

\subsection{Is There a Single Overall Best Solution?}
\reviewerfix{As the evaluation section makes clear,} there is no universally good library that achieves fast build throughput, fast query throughput for all query sizes and uses close to linear memory. However, Boost, CGAL and R-tree offer some of the best performance at large and small scales and have the advantage of being able to support both points and boxes. Of the three libraries, Boost offers the best overall performance in terms of good build and query times (and has the lowest memory usage), while CGAL offers the best build times and R-tree offers the best query times. There is therefore no single best solution, but rather the answer will depend on the problem scale, memory availability, the number of queries to be performed and the fraction of stored data that is expected to match the queries. In addition, as discussed in the next section, there are circumstances under which the best library to use is no library.

\subsection{Under What Circumstances Does a Brute Force Solution Offer the Best Performance?}
The results demonstrate that if many of the queries will retrieve 10\% or more of the stored mesh data, and if there is a small amount of data per process (e.g., less than 10 million points or elements), then one should strongly consider using a brute force solution. The libraries evaluated in this paper offer the best performance when the search space can be substantially reduced (pruning branches) and pay a large performance penalty for the $\log n$ tree traversals when a large portion of the tree must be searched. In addition, as discussed previously, all of the libraries require significantly more memory than the raw data requires, and therefore the brute force solution should be used if users are severely memory constrained.

\subsection{Future Work}
We leave as future work evaluating these libraries for dynamic use (e.g., if adaptive mesh refinement is used) and for mesh elements that are more complex shapes. 
\reviewerfix{
We also plan to assess the portability and software design choices of the libraries, and to evaluate how the use of different architectures affects the evaluation results. Finally, we plan to evaluate libraries for languages other than C and C++.}

%% file: acknowledgements.tex
\section*{Acknowledgements}
This paper describes objective technical results and analysis. Any subjective views or opinions that might be expressed in the paper do not necessarily represent the views of the U.S. Department of Energy or the United States Government.

Sandia National Laboratories is a multimission laboratory managed and operated by National Technology and Engineering Solutions of Sandia, LLC, a wholly owned subsidiary of Honeywell International, Inc., for the U.S. Department of Energy's National Nuclear Security Administration under contract DE-NA0003525.

Margaret Lawson acknowledges support from the United States Department of Energy through the Computational Sciences Graduate Fellowship (DOE CSGF) under grant number: DE- SC0020347.

This work was supported in part by the State of Illinois.
\newpage